\documentclass[10pt,floatfix,superscriptaddress]{revtex4}
\usepackage{amsfonts,amsthm,latexsym}
\usepackage{graphicx}
\usepackage{bm}
\usepackage{color}
\usepackage{soul} 
\usepackage{slashed}
\usepackage{mathrsfs}
\usepackage{latexsym}
\usepackage{longtable}
\usepackage{graphicx}
\usepackage{epsfig}
\usepackage{natbib}
\usepackage{amsmath}
\usepackage{bbm}
\usepackage{amssymb}
\usepackage{color}
\usepackage{hyperref}
\usepackage{url}
\usepackage{multirow}
\usepackage[utf8]{inputenc}
\usepackage{ulem}
\usepackage{soul}
\usepackage{subfigure}
\usepackage{scalerel}

\begin{document}

\author{Germ{\'a}n Malfatti}\email{gmalfatti@fcaglp.unlp.edu.ar}
  \affiliation{CONICET, Godoy Cruz 2290, Buenos Aires (1425), Argentina.}
  \affiliation{Grupo de Gravitaci\'on,
  Astrof\'isica y Cosmolog\'ia,\\ Facultad de Ciencias
  Astron{\'o}micas y Geof{\'i}sicas, Universidad Nacional de La
  Plata,\\ Paseo del Bosque S/N, La Plata (1900), Argentina.}

\author{Milva G. Orsaria} \email{morsaria@fcaglp.unlp.edu.ar}
\affiliation{CONICET, Godoy Cruz 2290, Buenos Aires (1425), Argentina.}
\affiliation{Grupo de Gravitaci\'on, Astrof\'isica y
  Cosmolog\'ia,\\ Facultad de Ciencias Astron{\'o}micas y
  Geof{\'i}sicas, Universidad Nacional de La Plata,\\ Paseo del Bosque
  S/N, La Plata (1900), Argentina.}

  \author{Gustavo A. Contrera} \email{contrera@fisica.unlp.edu.ar}
  \affiliation{CONICET, Godoy Cruz 2290, Buenos Aires (1425),
    Argentina.}  \affiliation{IFLP, UNLP, CONICET, Facultad de
    Ciencias Exactas, Diagonal 113 entre 63 y 64, La Plata (1900), Argentina.}
    \affiliation{Department of Physics, San Diego State University, 5500
  Campanile Drive, San Diego, CA 92182, USA.}

\author{Fridolin Weber} \email{fweber@sdsu.edu}
\affiliation{Department of Physics, San Diego State University, 5500
  Campanile Drive, San Diego, CA 92182, USA.}  \affiliation{Center for
  Astrophysics and Space Sciences, University of California,\\ San
  Diego, La Jolla, CA 92093, USA.}

\author{Ignacio F. Ranea-Sandoval} \email{iranea@fcaglp.unlp.edu.ar}
\affiliation{CONICET, Godoy Cruz 2290, Buenos Aires (1425), Argentina.}
\affiliation{Grupo de Gravitaci\'on, Astrof\'isica y
  Cosmolog\'ia,\\ Facultad de Ciencias Astron{\'o}micas y
  Geof{\'i}sicas, Universidad Nacional de La Plata,\\ Paseo del Bosque
  S/N, La Plata (1900), Argentina.}

\title{Hot quark matter and (proto-) neutron stars}

\begin{abstract}
In part one of this paper, we use a non-local extension of the
3-flavor Polyakov-Nambu-Jona-Lasinio model, which takes into account
flavor-mixing, momentum dependent quark masses, and vector
interactions among quarks, to investigate the possible existence of a
spinodal region (determined by the vanishing of the speed of sound) in
the QCD phase diagram and determine the temperature and chemical
potential of the critical end point.  In part two of the paper, we
investigate the quark-hadron composition of baryonic matter at zero as
well as non-zero temperature.  This is of great topical interest for
the analysis and interpretation of neutron star merger events such as
GW170817. With this in mind, we determine the composition of
proto-neutron star matter for entropies and lepton fractions that are
typical of such matter. These compositions are used to delineate the
evolution of proto-neutron stars to neutron stars in the baryon-mass
versus gravitational-mass diagram.  The hot stellar models turn out to
contain significant fractions of hyperons and $\Delta$-isobars but no
deconfined quarks. The latter, are found to exist only in cold neutron stars.
\end{abstract}



\maketitle

\section{Introduction}

Exploring the thermodynamic behavior of the quark-gluon plasma and its
associated equation of state (EoS) has become one of the forefront
areas of modern physics. The properties of such matter are being
probed with the Relativistic Heavy Ion Collider (RHIC) at BNL and the
Large Hadron Collider (LHC) at CERN, and great advances in our
understanding of such matter are expected from the next generation of
high density experiments at the Facility for Antiproton and Ion
Research (FAIR at GSI) \cite{CBMbook,FAIR-GSI}, the Nuclotron-bases
Ion Collider fAcility (NICA at JINR) \cite{NICAWP:2016,NICA-JINR}, the
Japan Proton Accelerator Research Complex (J-PARC at Tokai campus of
JAEA) \cite{J-PARC}, the Super Proton Synchrotron (SPS at CERN)
\cite{SPS-CERN} and the Beam Energy Scan (BES at BNL) \cite{BES-BNL}.

Depending on temperature $T$, and baryon chemical potential $\mu$, the
deconfined phase of quarks and gluons is believed to exist at two
extreme regions in the phase diagram of quantum chromodynamics
(QCD). The first regime corresponds to $T>> \mu$, which was the case
in the early Universe where the temperature was hundreds of MeV but
the net baryon number density was very low. Secondly, it is theorized
that quark deconfinement occurs also at low temperatures but very high
chemical potential, $T << \mu$, that is, at conditions which exist in
the inner cores of (proto-) neutron stars \cite{Roark2018}. Portions of the phase
diagram lying between these two extreme physical regimes can be probed
with relativistic collision experiments.

Effective field-theoretical models such as the Nambu-Jona-Lasinio
model and its extensions \cite{Buballa:2005,Fukushima:2008wg,
  Contrera:2007wu, Contrera:2010, Carlomagno2018} as well as lattice
QCD (LQCD) calculations
\cite{Philipsen:2012nu,Borsanyi:2013bia,Bellwied:2015rza} predict a
smooth crossover of nuclear matter to quark matter in the low density
but high temperature regime of the phase diagram. On the other hand,
in the low temperature but high chemical potential regime the
hadron-quark phase transition is likely be of first-order
\cite{Fukushima:2010bq}. Some recent works
\cite{Most:2018eaw,Bauswein:2018bma} have investigated the occurrence
of a first order phase transition in neutron-star mergers.

Spinodal instabilities are characteristic features in systems which
exhibit first-order phase transitions. If present in the quark gluon
plasma, spinodal instabilities would lead to density fluctuations that
have a qualitative influence on the dynamical evolution of the system
density \cite{Randrup:2012}.  The fluctuations that lead to a spinodal
decomposition are long range and differ from local fluctuations that
give rise to nucleation, which occurs in the metastable region of the
phase diagram. Hence, if there is a first-order phase transition,
large density fluctuations can arise as a result of spinodal
instabilities.  The effects of spinodal instabilities in nuclear
collision simulations at NICA energy densities were studied in
\cite{Randrup:2016}. The spinodal region has also been analyzed for
two \cite{Sasaki:2007, Sasaki:2008} and three
\cite{PhysRevC.95.055203} flavor quark matter using the local
Nambu-Jona-Lasinio (NJL) model.

In this work, we investigate the hadron-quark phase transition in
superdense matter and the possible appearance of deconfined
quark matter in the cores of (proto-) neutron stars.  For the
description of quark matter we use the non-local SU(3) NJL model
coupled to the Polyakov loop (hereafter referred to a 3nPNJL
model). Vector interactions among quark are taken into account too.
The use of non-local interactions has been suggested as an improvement
of the standard NJL model.  Non-locality arises naturally from several
successful approaches to low-energy quark dynamics, such as one-gluon
exchange descriptions \cite{Contrera:2007wu,Contrera:2010}, the
instanton liquid model \cite{Schafer:1996wv}, and the Schwinger-Dyson
resummation techniques \cite{Roberts:2000aa}.

We calculate the metastable regions of the phase diagram and
investigate the possible existence of quark-hybrid stars assuming a
sharp hadron-quark phase transition. Hadronic matter is assumed to be
made of neutrons, protons, hyperons, and delta isobars.  The field
equations of these particles are solved for an improved
parametrization of the relativistic mean-field model with density
dependent coupling constants.  The density at which quark
deconfinement may occur in the cores of neutron stars is assumed to
be several times greater than the saturation density of ordinary
nuclear matter. The equations of state computed in this work fulfill
the $2 M_{\odot}$ mass constraints set by PSR J1614-2230 and PSR
J0348+0432
\cite{Demorest:2010bx,Lynch:2012vv,Antoniadis:2013pzd,Arzoumanian_2018}
as well as the radius constraints derived from the gravitational-wave
event GW170817 and its electromagnetic counterpart GRB 170817A
\cite{Bauswein:2017vtn,Fattoyev:2017jql,Raithel:2018ncd,Most:2018hfd,Annala:2017llu}.

The article is organized as follows. In Section \ref{3nPNJL}, we
provide a description of the 3nPNJL model and its parameterizations,
pointing out some of the reasons for using a non-local model instead
of the local one. We analyze the structure of the phase diagram as
predicted by the 3nPNJL model and explore the occurrence of a spinodal
region, which is determined by the vanishing of the speed of
sound. Section \ref{hyper} is devoted to the description of hadronic
matter. In Section \ref{astro}, we discuss the transition of hadronic
matter to quark matter and present the properties of neutron stars
computed for the EoS of this work. The study of several selected
stages in the evolution of proto-neutron stars to neutron stars is
given in Section \ref{sec:apns}. Finally, Section
    \ref{summary} provides a summary of the results and some
    conclusions.

\section{Quark matter at finite temperature}
\label{3nPNJL}
\subsection{The local vs the non-local NJL model}

The standard NJL model is based on an effective lagrangian of relativistic
fermions interacting through local fermion-fermion couplings. Because
of the local nature of the interaction, the Schwinger-Dyson and
Bethe-Salpeter equations become relatively simple.  However, one of
the drawbacks of the model is that it is non-renormalizable.  The
problems of ultra-violet divergences for this model can be fixed by
using non-local rather than local interactions. Furthermore, the local
NJL model works with an artificial momentum-space cutoff of $\Lambda
\sim 0.6-0.7$~GeV, which is turned off at high momenta. Thus, the
applicability of this model is restricted to energy and momentum scales
(temperatures, chemical potentials) that are small compared to
$\Lambda$. Connections to the running QCD coupling constant and the
established high-momentum, high-temperature behavior governed by
perturbative QCD are therefore ruled out right from the start.

In this work we will consider the 3nPNJL model, which includes vector
interactions among quarks. Non-local extensions of the NJL model are
designed to remove the deficiencies of the local model, while, at the
same time, the non-local interactions regularize the model in such a
way that the basic features of a relativistic quark matter system,
like chiral symmetry breaking and the formation of bound stages in the
low-energy limit, can be properly described (see \cite{Contrera:2010}
and references therein).  In addition, the range (in momentum space)
of the non-locality provides a natural cutoff that falls off at high
densities, which makes the model more appropriate for the description
of quark matter than the local NJL model, even in the perturbative
regime.

\subsection{The 3nPNJL model}

To study the QCD phase diagram and the EoS with the 3nPNJL model, we
start from the lagrangian
\begin{eqnarray}\label{qm:lagrangian}
\mathcal{L}(x) &=& \bar \psi(x)(-i \slashed{D} + \hat{m} )\psi(x) +
\frac{G_V}{2}j_{a}^{\mu}(x)j_{a}^{\mu}(x) \nonumber \\ &-&
\frac{G_S}{2}\big[j_{a}^s(x)j_{a}^s(x) + j_{a}^p(x)j_{a}^p(x)\big] +
     {\cal U}\,[{\cal A}(x)] \\ &-& \frac{H}{4}A_{abc}\big[
       j_{a}^s(x)j_{b}^s(x)j_{c}^s(x) -
       3j_{a}^s(x)j_{b}^p(x)j_{c}^p(x) \big], \nonumber
\end{eqnarray}
which accounts for scalar as well as vector interactions among quarks.
The quantity ${\cal U}$ is an effective potential which accounts for
Polyakov loop dynamics, and the last term denotes the 'tHooft term
which is responsible for flavor mixing.  The quantities $\psi$ denotes
the light quark fields, $\psi \equiv (u\; d\; s)^T$, and $\hat m =
{\rm diag}(m_u, m_d, m_s)$ stands for the current quark mass
matrix. For simplicity we consider the isospin symmetric limit where
$m_u = m_d$.

Regarding the interaction terms, the scalar ($s$), pseudo scalar ($p$)
and vector ($\mu$) interaction currents are respectively given  by
\begin{eqnarray}
j_{a}^s(x) &=& \int d^4z \tilde{R}(z)\bar \psi \left(x + \frac{z}{2}
\right)\lambda_a \psi \left(x - \frac{z}{2} \right) \, , \nonumber
\\ j_{a}^p(x) &=& \int d^4z \tilde{R}(z)\bar \psi \left(x +
\frac{z}{2} \right) i\lambda_a \gamma^5 \psi \left(x - \frac{z}{2}
\right) \, , \\ j_{a}^{\mu}(x) &=& \int d^4z \tilde{R}(z)\bar
\psi \left(x + \frac{z}{2} \right) \lambda_a \gamma^{\mu}\psi \left(x
- \frac{z}{2} \right) \, ,\nonumber
\label{currents}
\end{eqnarray}
where $\tilde R$ is the Gaussian form factor whose Fourier transform
is given by $R(p) = \mathrm{exp}(-p^2/\Lambda^2)$, with $\Lambda$ being a
parameter that sets the range of non-locality in momentum space. The
matrices $\lambda_a$, with $a=0,..,8$, are the standard
Gell-Mann $3\times 3$ matrices (generators of SU(3)) and
$\lambda_0=\sqrt{2/3}\;I_{3\times 3}$. The constants $A_{abc}$
in the 'tHooft term are defined by
\begin{equation}
 A_{abc} =\frac{1}{3!} \epsilon_{ijk}\epsilon_{mnl} (\lambda_a )_{im}
 (\lambda_b )_{jn} (\lambda_c )_{kl }. \,
\end{equation}

The interaction between fermions and SU(3) color gauge fields
$G^a_\mu$ is described by the covariant derivative in the fermion
kinetic term, i.e., $D_\mu\equiv \partial_\mu - i{\cal A}_\mu$, where ${\cal
  A}_\mu$ will be defined, as usual, assuming that the quarks move in
a constant background field $A_4 = i A_0 = i g\,\delta_{\mu 0}\,
G^\mu_a \lambda^a/2$.

The partition function associated with the effective action $S_E=\int
d^4x \mathcal{L}(x)$ can be bosonized in the usual way introducing the
scalar, pseudoscalar and vector meson fields $\sigma_a(x)$,
$\pi_a(x)$, and $\theta_a$, respectively, together with auxiliary
fields $S_a(x)$, $P_a(x)$ and $V_a(x)$. To deal with these auxiliary
fields we follow the standard stationary phase approximation, which
provides a set of equations that relate them to the meson fields (the
procedure is similar to that described in
Refs. \cite{Scarpettini:2003fj,Carlomagno:2013}).  We consider the
mean field approximation (MFA), keeping only the nonzero vacuum
expectation values of the bosonic fields $\bar\sigma_a$ and
$\bar\theta_a$ and assuming that pseudoscalar mean field values
vanish, owing to parity conservation. Note that due to color charge
conservation only $\bar\sigma_{a=0,3,8}$ and $\bar\theta_{a=0,3,8}$
can be different from zero. In addition, $\bar\sigma_{3}$ also
vanishes in the isospin limit. It is therefore convenient to
transform the neutral fields $\bar\sigma_{a}$,
$\bar\theta_a$, $\bar S_a$ and $\bar V_a$ to a flavor basis
$(f=u,d,s)$ and to compute $\bar\sigma_{f}$, $\bar\theta_f$, $\bar
S_f$, and $\bar V_f$, as described in Ref. \cite{Contrera:2009hk}.

After bosonization of the effective action, the
regularized grand canonical potential in the mean-field approximation
(see Ref. \cite{GomezDumm:2004sr} for details in the regularization
procedure) follows as
\begin{equation}
\Omega = \Omega^{\rm reg} + \Omega^{\rm free} + \Omega^0
+\mathcal{U}(\Phi ,T)\, ,
\label{eq:Omega}
\end{equation}
where $\Omega^0$ is defined by the condition that $\Omega$ vanishes at
$T=\mu=0$. The effective potential ${\cal{U}}(\Phi ,T)$ can be fitted
by taking into account group theoretical constraints together with
lattice results, from which one can estimate the temperature
dependence. Following Ref.~\cite{Roessner:2006xn}, we take
\begin{eqnarray}
{\cal{U}}(\Phi ,T) &=& \left[-\,\frac{1}{2}\, a(T, T_0)\,\Phi^2 \;\right. \\
&+& \left.\;b(T, T_0)\, \ln(1
- 6\, \Phi^2 + 8\, \Phi^3 - 3\, \Phi^4)\right] T^4 \ ,\nonumber
\label{effpot}
\end{eqnarray}
with the definitions of $a(T, T_0)$ and $b(T, T_0)$ given in
Ref.~\cite{Roessner:2006xn}.  The parameter $T_0 = 195$ MeV is fixed
to reproduce LQCD results for the critical temperature (
\cite{Carlomagno2018,Carlomagno:2013} and references therein). Owing
to the charge conjugation properties of the QCD lagrangian, the mean
field traced Polyakov loop field $\Phi$, which serves as an order
parameter of confinement, is expected to be a real quantity
\cite{Contrera:2010}. Assuming that $\phi_3$ and $\phi_8$ are
real-valued, this implies that $\phi_8 = 0$. Then, $\Phi \equiv
\frac{1}{3} {\rm Tr}\, \exp( i \phi_c/T) = [ 2 \cos(\phi_3/T) + 1
]/3$, where the trace it to be taken with respect to the color
indices.  The color background fields $ \phi_c$ are $\phi_r = -\phi_g
= \phi_3$ and $\phi_b = 0$, thus $\phi_c = c \phi_3$ with $c =
\{-1,0,1\}$.

To study hot and dense quark matter we extend the bosonized effective
action to finite temperature using the Matsubara formalism. Thus, the
quantities $\Omega^{\rm reg}$ and $\Omega^{\rm free}$ in
Eq.\ (\ref{eq:Omega}) are given by
\begin{eqnarray}
\Omega^{\rm reg} = &-&2\,T
\sum_{f,c}\int\frac{p\,dp^3}{(2\,\pi)^3}\Bigg\{ 2\,
\sum_{n=0}^{\infty} \, \log \left[\frac{q_{fnc}^2 +
    M_{f}^2(w_{fnc}^2)}{w_{fnc}^2 + m_f^2}\right]\Bigg\} \nonumber
\\ &-&\frac{1}{2}\Bigg[ \sum_f \left(\bar\sigma_f \bar S_f +
  \frac{G_S}{2}\bar S_f^2 + \bar\theta_f \bar V_f - \frac{G_V}{2}\bar
  V_f^2\right) \nonumber \\ &+& \frac{H}{2}\bar S_u \bar S_d \bar
  S_s\Bigg], \nonumber \\ \Omega^{\rm free} = &-&2\,T
\sum_{f,c}\int\frac{p\, dp^3}{(2\,\pi)^3}\Bigg[\log \left(1+
  e^{-\frac{E_f - \mu_f - i \phi_c}{T}}\right) \nonumber\\ &+& \log
  \left(1+ e^{-\frac{E_f + \mu_f + i \phi_c}{T}}\right) \Bigg] \, ,
\end{eqnarray}
where $E_f=\sqrt{\vec p^{\;2} + m_f^2}$, $w_{fnc}^2 = (w_n - i\mu_f +
\phi_c)^2 + \vec p^{\;2}$, and $w_n$ denote the Matsubara
frequencies.  The shifted momentum becomes $q_{fnc}^2 = q_{0fnc}^2 +
\vec p\,^2$ with the zero-component given by $q_{0fnc}^2 = (w_n -
i[\mu_f - \bar\theta_f R(w_{fnc}^2)]+ \phi_c)^2$. The sums over flavor
and color indices run over $f = (u, d, s)$ and $c = ( r, g, b)$,
respectively. The momentum dependent constituent quark masses are
given by \newline $M_{f}(w_{fnc}^2) \ = \ m_f\, + \, \bar\sigma_f\,
R(w_{fnc}^2)$. Note that in the isospin limit $\bar \sigma_u = \bar
\sigma_d$, thus we have $M_{u} = M_{d}$. The mean-field values of the
auxiliary fields,
\begin{eqnarray}
 \bar S_f = -16T\sum_{c} \int \frac{p\, dp^3}{(2\,\pi)^3}\sum_{n=0}^{\infty}
 \,\,\frac{ M_{f}(w_{fnc}^2) R(w_{fnc}^2)}{q_{fnc}^2 +
   M_{f}^2(w_{fnc}^2)},\nonumber \\ \bar V_f = -16T\sum_{c} \int
 \frac{p\, dp^3}{(2\,\pi)^3}\sum_{n=0}^{\infty} \,\,\frac{ i\,q_{0fnc}
   R(w_{fnc}^2)}{q_{fnc}^2 + M_{f}^2(w_{fnc}^2)} \, ,
 \label{eq:Aux}
\end{eqnarray}
are obtained by minimizing the thermodynamic potential with respect to
the mean-field values $\bar \sigma_f$ and $\bar\theta_f$,
respectively.  Minimizing $\Omega$ with respect to the mean-field
values and the Polyakov-loop color field $\phi_3$ leads to a system of
coupled non-linear equations that can be solved numerically for the
mean-field values in Eqs.\ (\ref{eq:Omega}) and (\ref{eq:Aux}). From
the grand canonical potential $\Omega$ the system's energy density,
$\epsilon$, pressure, $P$, and quark number density, $n_q$ follow as
\begin{eqnarray}
 \epsilon = - P + T S + \sum_f \mu_f n_f \, ,  \nonumber\\
 P = - \Omega \, ,  \;\;\;\;\;\; n_q = \sum_f n_f \, ,
 \label{qm:themodynamics}
\end{eqnarray}
with $S = \frac{\partial P}{\partial T}$ and $n_f = \frac{\partial P}{\partial \mu_f}$.

To regulate the non-local interactions we use the Gaussian form factor
$R(w_{fnc}^2) = \mathrm{exp}(-{w_{fnc}^2}/{\Lambda^2})$.  The argument
of the form factor, $w_{fnc}^2$, is not shifted by the vector
interaction because the regulator is inserted as a distribution
function in the lagrangian before taking the mean values of the
fields. The up ($m_u$) and down ($m_d$) current quarks masses and the
coupling constants $G_S$, $H$, and $\Lambda$ are chosen so as to
reproduce the phenomenological values of the pion decay constant,
$f_\pi=92.4$ MeV, and the meson masses $m_{\pi}=139.0$ MeV, $m_K=495$
MeV, $m_{\eta'}=958$ MeV
\cite{Contrera:2007wu,Contrera:2009hk,JPGreview}, leading to $m_u =
m_d = 3.63$ MeV, $\Lambda = 1071.38$ MeV, $G_s \Lambda^2 = 10.78$, and
$H \Lambda^5 = -353.29$. The strange quark current mass is set to an
updated phenomenological value of $m_s = 95.00$ MeV, and $m_s/m_u
\simeq 26$ is in agreement with the latest data provided by the
Particle Data Group \cite{Tanabashi:2018oca}.

The vector interaction coupling constant $G_V$ is usually expressed in
terms of the scalar coupling constant, $G_S$. In what follows, we
introduce the quantity $\mathrm{\zeta_v} \equiv G_V/G_S$ to denote the
vector-to-scalar interaction strength. As it is customary, we treat
$G_V$ as a free parameter, due the uncertainty in its theoretical
predictions \cite{Contrera:2012wj}. Different values for
$\mathrm{\zeta_v}$ will be chosen in the next sections to show the
effect of the vector interaction on the properties of quark matter.
\begin{figure}[ht]
\begin{center}
\includegraphics[width=0.48\textwidth]{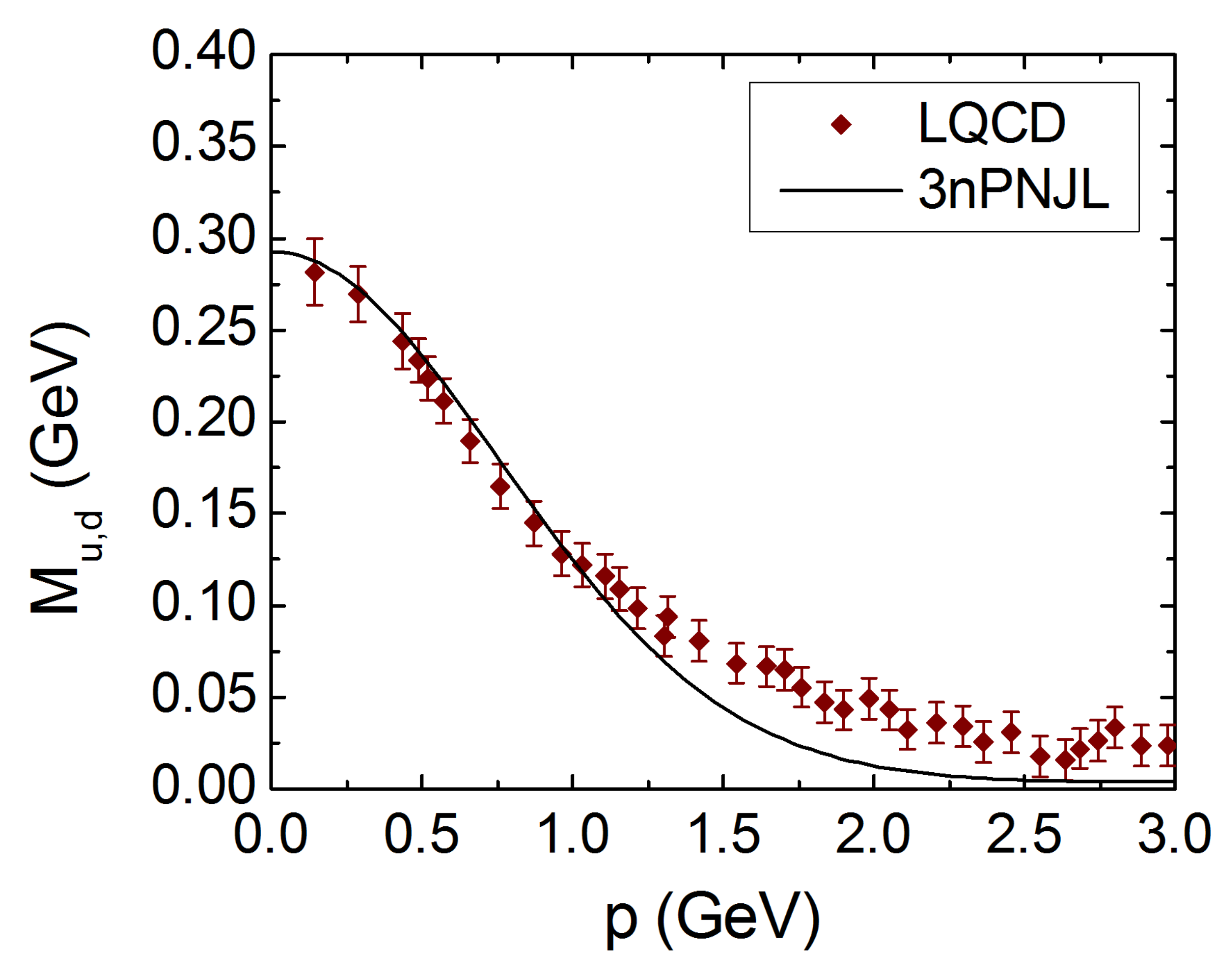}
\caption{(Color online) Dependence of the dynamical masses of light
  ($u$ and $d$) quarks on momentum $p$ for the parametrization used in
  this work. (solid line). The diamond shaped symbols show the results
  of LQCD calculations for $N_f = 2 + 1$ quark flavors extracted from
  \cite{Parapilly:2006}.}
 \label{Massa_LQCD}
\end{center}
\end{figure}
The form factor $R(p)$, defined in Eq.\ (\ref{currents}) and the given
parameters, guarantee a rapid ultra-violet convergence of the loop
integrals. As can be seen in Fig.\ \ref{Massa_LQCD}, the functional
form of the form factor is chosen such that the momentum dependence of
the dynamical quark mass is reproduced by the light quarks masses obtained
in LQCD calculations \cite{Parapilly:2006}.

\subsection{Spinodal decomposition and the QCD phase diagram}

As indicated by extensions of LQCD
to finite chemical potentials for finite quark masses
\cite{Philipsen:2012nu,Borsanyi:2013bia,Bellwied:2015rza}, there
should be a crossover phase transition in the QCD phase diagram at low
chemical potential. In addition, the study of some extrapolations to
the continuum limit for $2+1$ quark flavors
\cite{Bazavov:2011nk,Aoki:2009sc,Bazavov:2016uvm} give a critical
temperature of around $T_c(0)\simeq155$ MeV.

\begin{figure}[ht]
\begin{center}
\includegraphics[width=0.49\textwidth]{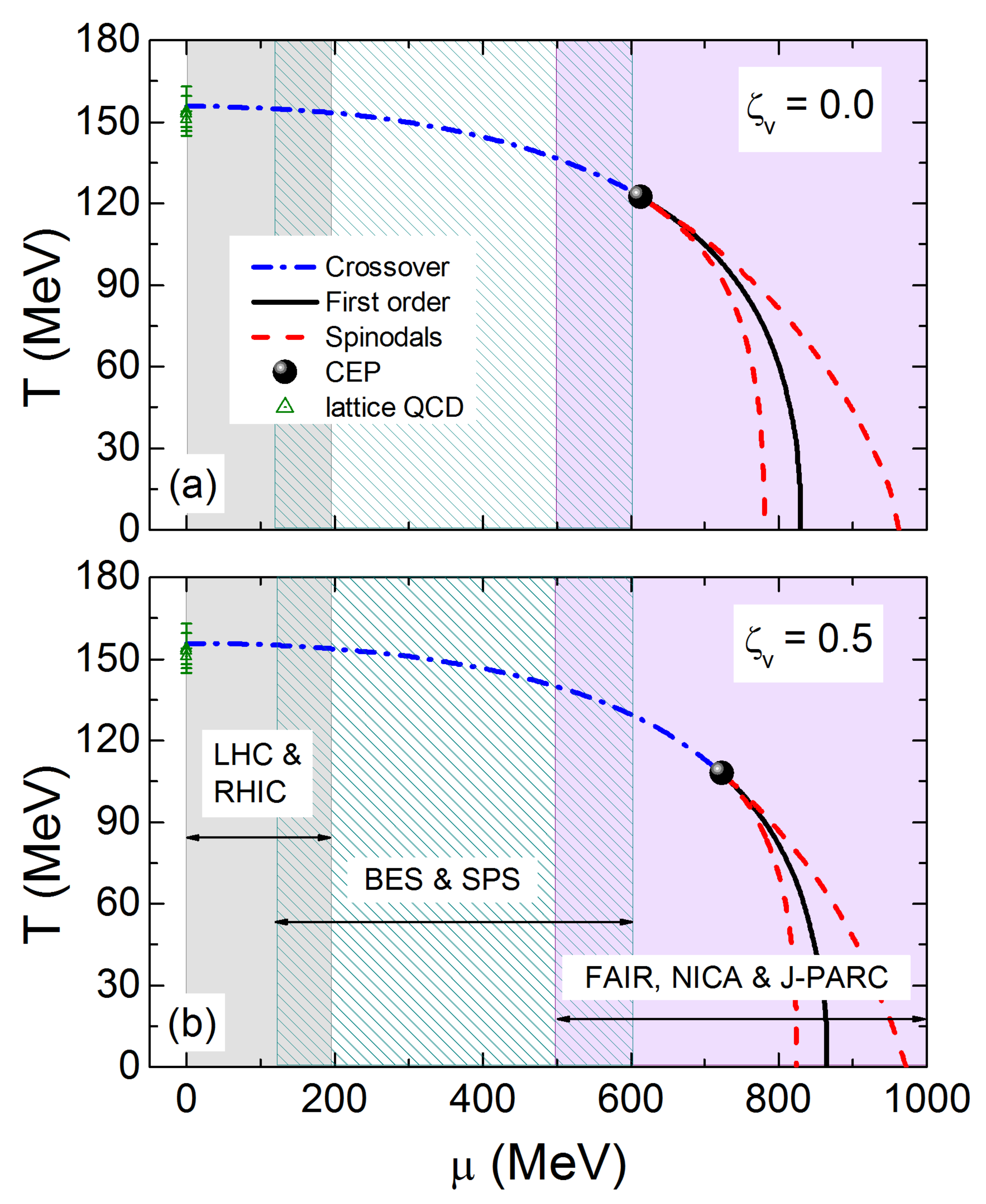}
\caption{(Color online) Temperature, $T$, versus baryon chemical
  potential, $\mu$, for 3-flavor quark matter without (a) and
  with (b) vector interactions. LQCD results
  \cite{Bazavov:2011nk,Aoki:2009sc,Bazavov:2016uvm}
  are marked. The crossover phase transition is shown by the
  dot-dashed blue line.  The dashed red line shows the spinodals, the
  solid black lines marks the first-order phase transition. The
  location of the critical endpoint (CEP) is shown by the solid dot.}
  \label{phase_sin_ch_n}
\end{center}
\end{figure}

At large chemical potentials but low temperatures, on the other hand,
a first-order phase transition is expected based on phenomenological
studies of quark matter (see, for example, \cite{Fukushima:2010bq},
and references therein).  This suggests that there should be a
second-order phase transition critical end point (CEP) at some
critical temperature and critical chemical potential, where the
different phase transitions meet. The location of the CEP and the
signatures of the first-order phase transition are being investigated
in the new experimental facilities as NICA, FAIR and J-PARC, while the
intermediate density (crossover) region is the target of the renewed
facilities BES and SPS at RHIC and CERN, respectively.  These regions
are shown in Fig.\ \ref{phase_sin_ch_n}.

 It is worth noting that according to LQCD simulations at finite
  temperature and zero chemical potential, chiral symmetry restoration
  occurs approximately simultaneously with quark deconfinement.
  The restoration of such symmetry and the consequent melting of the chiral condensate, defined in our model as $\langle\bar\psi_f\psi_f\rangle\,=\,\partial_{m_{f}}\,\Omega$, takes
  place already in the hadronic phase by parity doubling
  \cite{Aarts2017}, which is signaled by a mass degeneracy of
  hadronic chiral partner states. A model that restores chiral
  symmetry in the hadronic phase by lifting the mass splitting between
  chiral partner states before quark deconfinement sets in has
  recently been studied in Ref. \cite{Marczenko2018}.

\begin{figure}[ht]
\begin{center}
\includegraphics[width=.47\textwidth]{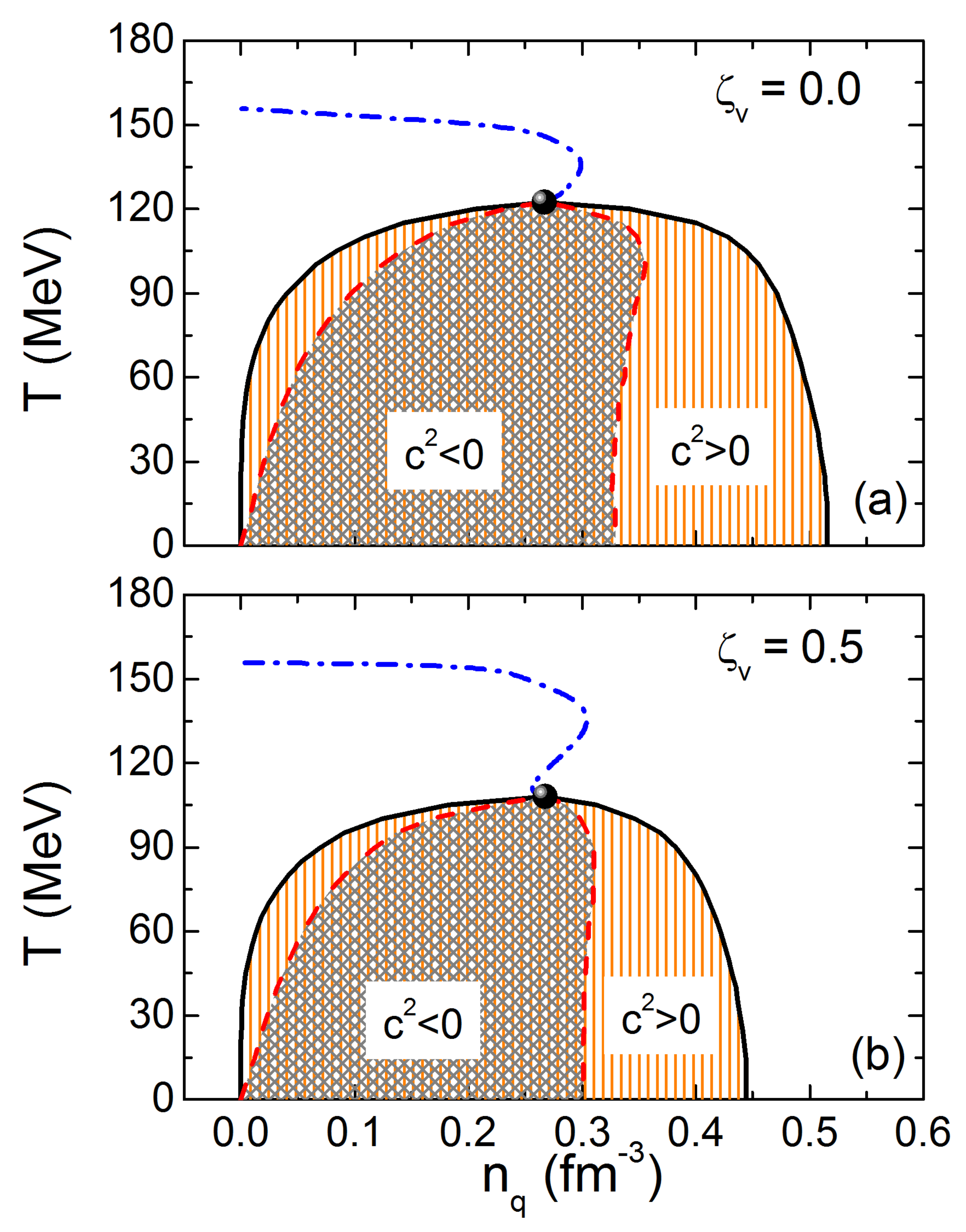}
\caption{(Color online) Temperature, $T$, versus quark number density,
  $n_q$, for 3-flavor quark matter without (a) and with (b)
  vector interactions. The crossover and the first order phase
  transitions are shown by the dot-dashed blue and solid black lines,
  respectively. The critical endpoint (CEP) is marked with a solid
  black dot. Unstable ($c_{s}^2 < 0$) and metastable ($c_{s}^2 >0$)
  regions are highlighted.}
  \label{Spinodals_T-nq}
\end{center}
\end{figure}

In Figs. \ref{phase_sin_ch_n} and \ref{Spinodals_T-nq} we show the
phase diagram of quark matter computed with the 3nPNJL model
introduced in Sect.\ \ref{3nPNJL}. The baryon chemical potential is
given by $\mu=\sum_f \mu_f$. In the crossover region (blue dot-dashed
line) of Fig.\ \ref{phase_sin_ch_n}, the critical temperatures obtained
from LQCD results
\cite{Bazavov:2011nk,Aoki:2009sc,Bazavov:2016uvm} are marked by green
triangles.  In addition, we have indicated the regions explored by
the Beam Energy Scan (BES) of the STAR collaboration at RHIC and by
the ALICE Collaboration at the LHC.
The first-order phase transition is shown by a black solid line, and
the critical endpoint (CEP) is marked with a solid black dot. Finally,
the spinodal lines, marked by red dashed lines, show the limit of the
metastable regions which will be explained later. It is
    important to note that the phase diagram shown in
    Fig. \ref{phase_sin_ch_n} is for quark matter only.  This figure,
    therefore, should not be confused with the full QCD phase
    diagram. A recent discussion of the QCD phase diagram based on a
    hadronic model and a chiral quark model (which is simpler than the 3nPNJL
    model of this work) can be found in Ref.
\cite{TKlahn2017}.

In order to show the effects of vector interaction we have chosen the
values $\mathrm{\zeta_v} = 0.0$ and $\mathrm{\zeta_v} = 0.5$, the
latter being the standard value that follows from the Fierz
transformation of the interaction between quark color currents induced
by gluon exchange \cite{Zhang2009}.
\begin{figure}[ht]
\begin{center}
\includegraphics[width=0.45 \textwidth]{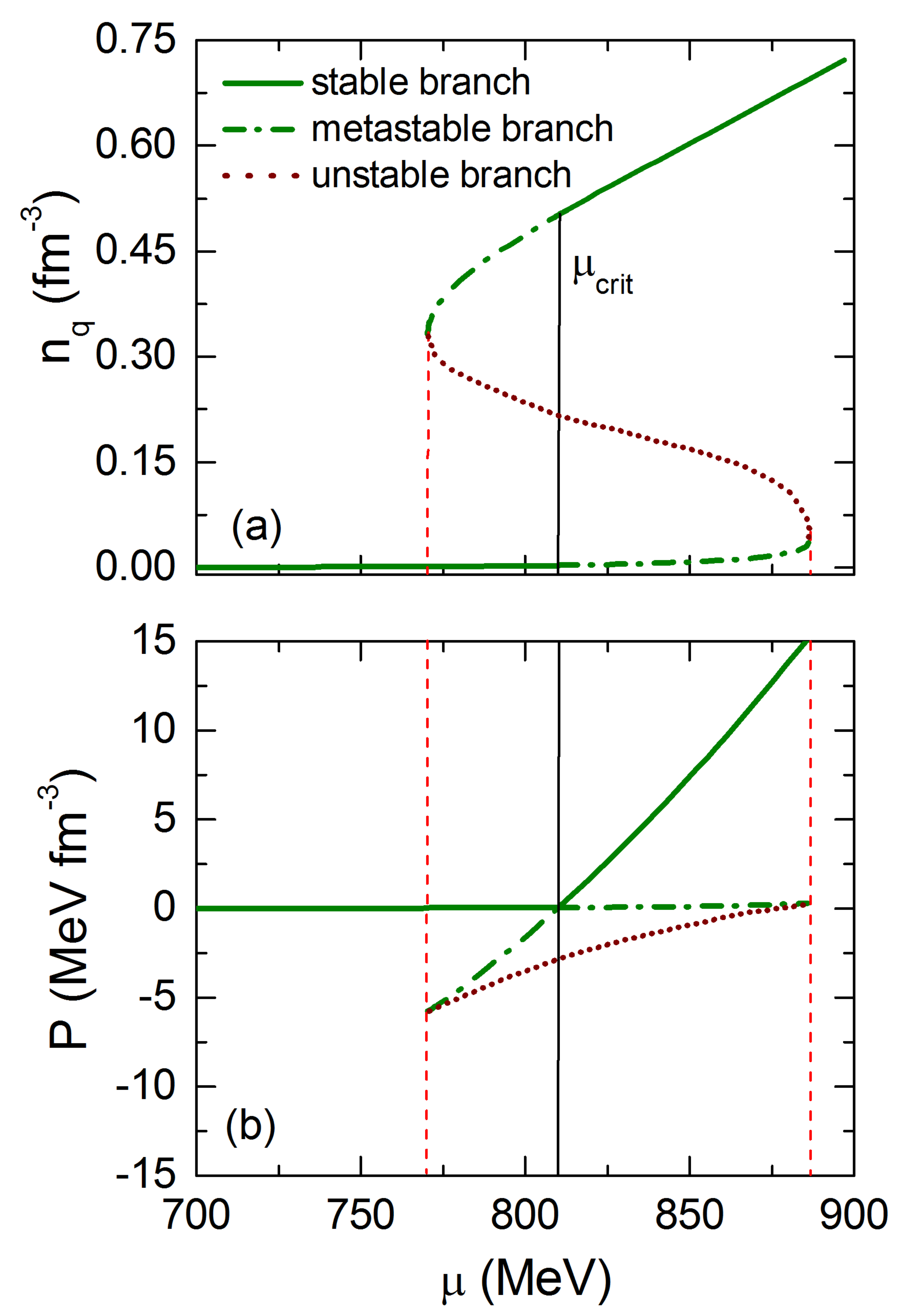}
\caption{(Color online) (a) Quark number density, $n_q$, and (b) pressure,
  $P$, as a function of baryon chemical potential,
  $\mu$, at $T = 50$ MeV. The dotted, solid and dot-dashed lines
  show unstable, stable, and metastable equilibrium, respectively.}
  \label{granpo_rho}
\end{center}
\end{figure}
Using the SU(3) version of the local PNJL model, it has been shown
\cite{Fukushima:2008is} that the inclusion of repulsive vector
interactions among quarks shrinks the first-order transition region by
moving the CEP to lower temperatures but higher densities, eventually
causing the CEP to vanish at high enough values of the vector coupling
constant $\mathrm{\zeta_v}$. However, the value chosen for
$\mathrm{\zeta_v}$ in this work allows for the existence of a CEP, in
agreement with the results of LQCD extrapolation techniques
\cite{Ratti_2018}. By comparing panels (a) and (b) in
Fig.\ \ref{phase_sin_ch_n}, it can be seen that for the 3nPNJL model
used in our work, the inclusion of vector interactions shifts the
first-order phase transition to higher chemical potentials and lower
temperatures. Finally, the results displayed in panels (a) and (b) in
Fig.\ \ref{Spinodals_T-nq} show that vector interactions tend to
shrink the regions (gray areas) where metastable quark matter exists.

The crossover phase transition is determined by the peaks of the
chiral susceptibility, as in \cite{Contrera:2007wu,Contrera:2010}.
The method of construction of the phase diagram in the ($T, n_q$)
plane for the first-order phase transition follows from
Fig.\ \ref{granpo_rho}. The dotted lines show unstable equilibrium,
solid and dot-dashed lines show stable and metastable equilibria,
respectively. The critical first order values ($T_{\rm crit}$,
$\mu_{\rm crit}$) used to construct the phase coexistence line in
panels (a) and (b) of Fig.\ \ref{phase_sin_ch_n} are defined by the
point where the zig-zag shaped branches of the pressure $P$ cross each
other.

The region where density fluctuations associated with the spinodals
occurs can be analyzed in term of the isothermal speed of sound, $c_s$
given by \cite{randrup:2009,randrup:2010}
\begin{eqnarray}
c_{s}^2 = \frac{n_q}{\epsilon + P} \left( \frac{\partial P}{\partial
  n_q} \right)_T \, .
\end{eqnarray}

The gray-shaded regions in Fig.\ \ref{Spinodals_T-nq} show
unstable regions in the phase diagram where $c_{s}^2 < 0$. These regions are
surrounded by metastable regions shown in orange where $c_{s}^2 >0$. The
dashed red curves show the spinodal lines determined by $c_{s}^2 = 0$,
while the blue dot-dashed and the solid black curves show the
crossover and first-order phase transitions, respectively.
\begin{figure}[ht]
\includegraphics[width=.45\textwidth]{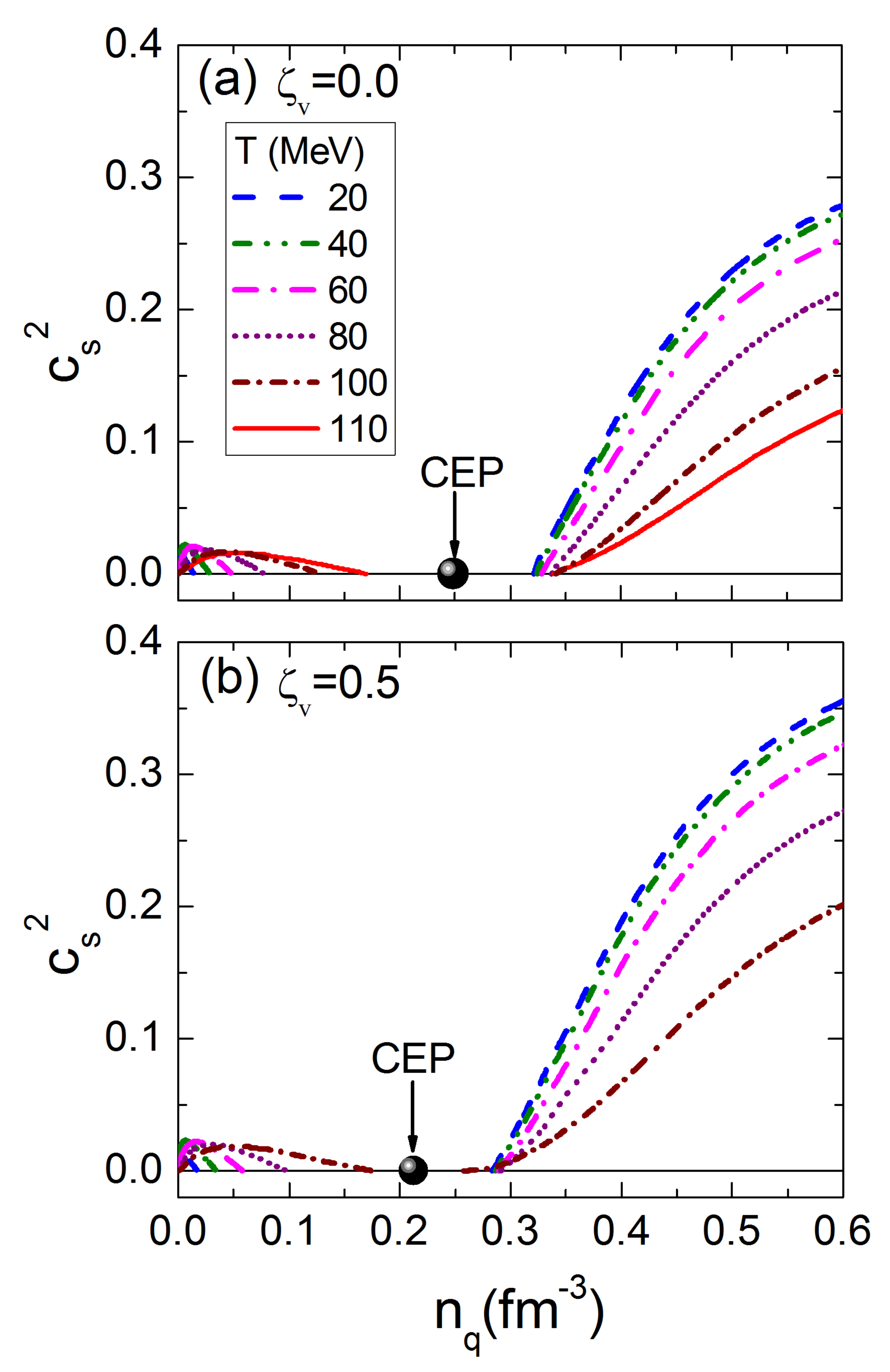}
\caption{(Color online) Isotherms of the square of the speed of sound,
  $c_s^2$, as a function of quark number density, $n_q$ without (a),
  and with (b) vector interactions among quarks. The solid dots
  indicate the location of the critical endpoints (CEP).}
\label{cs2}
\end{figure}
In the region where $c_{s}^2 < 0$, the ``compressibility'' $\kappa
\propto n_q(\frac{\partial P}{\partial n_q})_T$ \cite{Schmitt:2010pn})
is negative and the system responds to an increase in density by
enlarging any small density fluctuations. Since this region is not
stable, all the density fluctuations that normally occur in the zone
bounded by the isothermal spinodals will separate the system into
regions of low density and high density. The spinodal curves separate
unstable regions from metastable regions in Fig.\ \ref{Spinodals_T-nq}.
The right branch of the spinodal curve
shows the regions where an increase in density in the denser phase
does not cause any change in pressure. The left branch of the spinodal
shows the equivalent to this, but for the less dense phase. The
metastable region is bounded by the coexistence region and the
isothermal spinodal curve. It is in this region where density
fluctuations either grow through the aggregation of quark condensates
(left branch) or shrink because of the evaporation of these
condensates (right branch). It is worth noticing that if one wants to construct an equation of
state for deconfined quark matter, it is necessary to work with
chemical potentials that lie on the right-hand side of the spinodal
lines so that perturbations do not lead to the formation of mesons.

The behavior of $c_s^2$ (in units of the speed of light) as a function
of quark number density is shown in Fig.\ \ref{cs2} for different temperatures.
  Note that for the cases without vector interactions
$c_s^2$ is less than $1/3$, as suggested for weakly interacting quark
matter \cite{Bedaque:2014sqa}. Non-vanishing vector interactions among
quarks stiffen the EoS and the speed of sound increases to values
greater than $1/3$. (The region where $c_s^2 < 0$, which correspond to
the unstable region of the first-order phase transitions, has been
omitted in Fig.\ \ref{cs2}.)

\section{Hadronic matter at finite temperature}
\label{hyper}

In the most primitive conception, the matter in the core of a neutron
star is constituted from neutrons. At a slightly more accurate
representation, the cores consist of neutrons and protons whose
electric charge is balanced by leptons ($L=\{e^-,\mu^-\}$). Other
particles, like hyperons ($B=\{n,p,\Lambda,\Sigma,\Xi\}$) and the
$\Delta$-isobar, may be present if the Fermi energies of these
particles become large enough so that the existing baryon populations
can be rearranged and a lower energy state be reached.  To model this
hadronic phase, we make use of the density-dependent relativistic mean-field
(DDRMF) theory, in which the interactions between baryons are
described by the exchange of scalar ($\sigma$), vector ($\omega$), and
isovector ($\rho$) mesons. The lagrangian of this model is given by
\begin{eqnarray}
  \mathcal{L} &=& \sum_{B}\bar{\psi}_B \bigl[\gamma_\mu [i\partial^\mu
  - g_{\omega B}(n)
    \omega^\mu
    - g_{\rho B}(n) {\boldsymbol{\tau}} \cdot {\boldsymbol{\rho}}^\mu]
    \nonumber\\
     & -& [m_B - g_{\sigma B}(n)\sigma]
    \bigr] \psi_B + \frac{1}{2} (\partial_\mu \sigma\partial^\mu
  \sigma  - m_\sigma^2 \sigma^2)\\
  & -& \frac{1}{3} \tilde{b}_\sigma m_N [g_{\sigma N}(n)
    \sigma]^3 - \frac{1}{4} \tilde{c}_\sigma [g_{\sigma N}(n) \sigma]^4 -
  \frac{1}{4}\omega_{\mu\nu} \omega^{\mu\nu}\nonumber\\
 & +&\frac{1}{2}m_\omega^2\omega_\mu \omega^\mu + \frac{1}{2}m_\rho^2
  {\boldsymbol{\rho\,}}_\mu \cdot {\boldsymbol{\rho\,}}^\mu - \frac{1}{4}
  {\boldsymbol{\rho\,}}_{\mu\nu} \cdot {\boldsymbol{\rho\,}}^{\mu\nu} \, ,\nonumber
  \label{eq:Blag}
\end{eqnarray}
where $g_{\sigma B}(n)$, $g_{\omega B}(n)$ and $g_{\rho B}(n)$ are
density dependent meson-baryon coupling constants and $n = \sum_B n_B$
is the total baryon number density.  The density dependent coupling
constants are given by \cite{Typel:2018cap}
\begin{equation}
g_{i B}(n) = g_{i B}(n_0)\,\, a_{i}\,
\frac{1+b_{i}(\frac{n}{n_0}+d_{i})^{2}}{1+c_{i}(\frac{n}{n_0}+d_{i})^{2}}\, ,
\end{equation}
for $i=\sigma,\omega$ and
\begin{equation}
g_{\rho B}(n) = g_{\rho B}(n_0)\,\mathrm{exp}\left[\,-a_{\rho} \left(\frac{n}{n_0} -  1\right)\,\right]\, .
\end{equation}
 This choice of parametrization accounts for nuclear medium effects
 \cite{Fuchs:1995as}. The parameters $a_{i}$, $b_{i}$, $c_{i}$, and
 $d_{i}$ are fixed by the binding energies, charge and diffraction
 radii, spin-orbit splittings, and the neutron skin thickness of
 finite nuclei. Note that the density dependence of the meson-baryon
 couplings in the DD2 parametrization eliminates the need for
 non-linear self-interactions of the $\sigma$ meson. Therefore, the
 non-linear terms in the lagrangian given in Eq. (\ref{eq:Blag}) are
 considered only for the GM1L parametrization.

The meson-hyperon coupling constants have been determined following
the Nijmegen extended soft core (ESC08) model \cite{ECSO_SU3}. The
relative isovector meson-hyperon coupling constants were scaled with
the hyperon isospin and for the $\Delta$-isobar $x_{{\sigma}
  {\Delta}} = x_{{\omega} {\Delta}} = 1.1$ and $x_{{\rho} {\Delta}} =
1.0$, where $x_{ i H} = g_{i H}/g_{i N}$ was used (see
\cite{Spinella2017:thesis} for details).

In Table \ref{table:parametrizations}
we list the parameters of the DDRMF models used in this work.
Table
\ref{table:properties} shows the
saturation properties of the models, which are the nuclear saturation density,
$n_0$, energy per nucleon, $E_0$, nuclear incompressibility, $K_0$,
effective nucleon mass, $m^*/m_N$, asymmetry energy, $J$, slope of the asymmetry
energy, $L_0$, and the nucleon potential, $U_N$.

\begin{table}[htb]
\begin{center}
\begin{tabular}{|c|c|c|}
\hline 
$~~$Parameters$~~$ & $~~$GM1L$~~$
&DD2$~~$\\ \hline
$m_{\sigma}$  (GeV)    & 0.5500      & 0.5462     \\
$m_{\omega}$  (GeV)          &0.7830    & 0.7830  \\
$m_{\rho}$  (GeV)          & 0.7700        & 0.7630    \\
$g_{\sigma N}$             & 9.5722       & 10.6870    \\
$g_{\omega N}$            & 10.6180         & 13.3420     \\
$g_{\rho N}$            & 8.9830         & 3.6269   \\
$\tilde{b}_{\sigma}$         &0.0029                  & 0         \\
$\tilde{c}_{\sigma}$         &$- 0.0011$                 &   0       \\
$a_{\sigma}$         &0          &1.3576   \\
$b_{\sigma}$         & 0                  & 0.6344          \\
$c_{\sigma}$         & 0                 &  1.0054        \\
$d_{\sigma}$         & 0                 & 0.5758         \\
$a_{\omega}$         &0           &1.3697   \\
$b_{\omega}$         & 0               &0.4965          \\
$c_{\omega}$         & 0                 &  0.8177        \\
$d_{\omega}$         &0                  & 0.6384         \\
$a_{\rho}$         &0.3898           &0.5189   \\ \hline
\end{tabular}
  \caption{Parameters of the DDRMF parametrizations that lead to the
    properties of symmetric nuclear matter at saturation density given
    in Table \ref{table:properties}.}
\label{table:parametrizations}
\end{center}
\end{table}
\begin{table}[htb]
\begin{center}
\begin{tabular}{|c|c|c|}
\hline 
$~~$Saturation Properties$~~$ & $~~$GM1L$~~$
&DD2$~~$\\ \hline
$n_0$  (fm$^{-3}$)    & 0.153      & 0.149     \\
$E_0$  (MeV)          & $-16.30$    & $-16.02$  \\
$K_0$  (MeV)          & 300.0        & 242.7    \\
$m^*/m_N$             & 0.70       & 0.56    \\
$J$    (MeV)          & 32.5         & 32.8     \\
$L_0$  (MeV)          & 55.0         & 55.3    \\
$-U_N$ (MeV)        &65.5           &75.2   \\ \hline
\end{tabular}
  \caption{Properties of nuclear matter at saturation density computed
    for the DDRMF parametrizations GM1L
    \cite{Spinella2017:thesis,Spinella:2018bdq} and DD2
    \cite{Typel:2009sy}.}
\label{table:properties}
\end{center}
\end{table}

The meson mean-field equations following from Eq.\ (\ref{eq:Blag}) are given by
\begin{eqnarray}
m_{\sigma}^2 \bar{\sigma} &=& \sum_{B} g_{\sigma B}(n) n_B^s -
\tilde{b}_{\sigma} \, m_N\,g_{\sigma N}(n) (g_{\sigma N}(n)\bar{\sigma})^2
\nonumber\\ & & - \tilde{c}_{\sigma} \, g_{\sigma N}(n) \, (g_{\sigma N}(n)
\bar{\sigma})^3 \, , \nonumber\\ m_{\omega}^2 \bar{\omega} &=& \sum_{B}
g_{\omega B}(n) n_{B}\, , \\ m_{\rho}^2\bar{\rho} &=& \sum_{B}g_{\rho
  B}(n)I_{3B} n_{B} \, , \nonumber
\end{eqnarray}
where $I_{3B}$ is the 3-component of isospin and $n_{B}^s$ and $n_{B}$
are the scalar and particle number densities for each baryon $B$,
which are given by
\begin{eqnarray}
n_{B}^s&=& \gamma_B \int \frac{d^3p}{(2 \pi)^3} \left[f_{B-}(p) -
  f_{B+}(p)\right] \frac{m_B^*}{E_B^*}, \\ n_{B}&=& \gamma_B \int
\frac{d^3p}{(2 \pi)^3} \left[f_{B-}(p) - f_{B+}(p)\right] \, .
\end{eqnarray}
Here $f_{B\mp}$ denotes the Fermi-Dirac distribution function and
$E^*_B$ stands for the effective baryon energy given by
\begin{equation}
f_{B\mp}(p)=\frac{1}{\exp\left[\frac{E_B^*(p) \mp \mu_B^*}{T}\right] +
  1},\,\,\,\, E_B^*(p)=\sqrt{p^2 + m_B^{*2}}\, ,\nonumber
\end{equation}
where $\gamma_B=2J_B+1$ is the spin degeneration factor and $m_B^*=
m_B - g_{\sigma B}(n)\bar{\sigma}$ is the effective baryon
mass. We shall note at this point, that this model does not
    distinguish from parity in mass eigenstates. Because of that, the
    neutron mass is set to $m_N = 939.6$ MeV and that is the value
    that it takes when the background $\sigma$ field goes to zero.
    For a detailed explanation of a model that distinguish hadronic
    chiral partner states see \cite{Marczenko:2017huu}. The
effective chemical potential, $\mu_B^*$, is given by
\begin{equation}
\mu_B^* = \mu_B - g_{\omega B}(n) \bar{\omega} - g_{\rho B}(n)
\bar{\rho} I_{3B} - \widetilde{R} \, ,
\end{equation}
where $\widetilde{R}$ is the rearrangement term given by
\begin{eqnarray}
\widetilde{R} =\sum_B&&\left( \frac{\partial g_{\omega B}(n)}{\partial
  n} n_B \bar{\omega} + \frac{\partial g_{\rho B}(n)}{\partial n}
I_{3B} n_B \bar{\rho} \right. \\ &-& \left.\frac{\partial g_{\sigma
    B}(n)}{\partial n} n_B^s \bar{\sigma}\right) \, , \nonumber
\label{rear}
\end{eqnarray}
which is important for achieving thermodynamical consistency
\cite{Hofmann:2001}.  This term also contributes to the total baryonic
pressure of the matter,
\begin{eqnarray}
P &=& \sum_B \frac{\gamma_B}{3} \int \frac{d^3p}{(2 \pi)^3}
\frac{p^2}{E_B^*} [f_{B-}(p) + f_{B+}(p)] \nonumber\\ &-& \frac{1}{2}
m_{\sigma}^2 \bar{\sigma}^2 + \frac{1}{2} m_{\omega}^2 \bar{\omega}^2
+ \frac{1}{2} m_{\rho}^2 \bar{\rho}^2\\ &-& \frac{1}{3}
\tilde{b}_{\sigma} m_N (g_{\sigma N}(n) \bar{\sigma})^3 - \frac{1}{4}
\tilde{c}_{\sigma} (g_{\sigma N}(n) \bar{\sigma})^4 + n \widetilde{R}.
\nonumber
\label{HM:pressure}
\end{eqnarray}

The expression for the energy density, $\epsilon$, is determined by
the Gibbs relation given in Eq. (\ref{eq:EoS}).

\section{Neutron star matter and neutron stars}
\label{astro}

For the description of the matter inside of (proto-) neutron stars,
leptons must be also taken into account in both, the hadronic and the quark matter models.
They can be treated as free Fermi gases with the grand canonical potential given by
\begin{equation}
\Omega_L = - \sum_L \frac{\gamma_L}{3} \int \frac{d^3p}{(2 \pi)^3}
\frac{p^2}{E_L} [f_{L-}(p) + f_{L+}(p)] \, ,
\label{eq:leptons}
\end{equation}
with the lepton distribution function given by
\begin{eqnarray}
f_{L\mp}(p)&=&\frac{1}{\exp\left[\frac{E_L(p) \mp \mu_L}{T}\right] +
  1},\,\,\,\,\,\, E_L(p)=\sqrt{p^2 + m_L^{2}} \, . \nonumber
\end{eqnarray}
The lepton degeneracy factor is given by $\gamma_L=2$. The sum over
$L$ in Eq.\ (\ref{eq:leptons}) runs over $e^-$ and $\mu^-$ with masses
$m_L$ and, when correspond (see Sect. \ref{sect:PNS_evol}), massless neutrinos, $\nu_e$.

In addition, the composition of the matter in a neutron star is constrained by
charge neutrality and $\beta-$equilibrium. Electric charge and baryon
number are conserved. The conditions of electric charge neutrality and of
baryon number conservation lead to
\begin{equation}
\sum_B q_B\,n_B + \sum_L q_L\,n_L = 0 \, ,
\end{equation}
and
\begin{equation}
\sum_B \,n_B - n = 0 \, ,
\end{equation}
where the subscripts $B$ and $L$ stand for baryons and leptons
respectively, $q_i$ is the electric charge of these particles.

 The condition of chemical equilibrium reads
\begin{equation}
\mu_B = \mu_n - q_B (\mu_e - \mu_{\nu_e}) \, ,
\label{chem}
\end{equation}
where $\mu_n$, $\mu_e$ and $\mu_{\nu_e}$ are the neutron, electron and neutrino chemical
potentials, respectively. For the quark matter phase,
this condition is given by
\begin{equation}
\mu_f = \widetilde{\mu} - q_f (\mu_e - \mu_{\nu_e}) \, ,
\end{equation}
where $\mu_n$ is replaced by an average quark chemical potential
$\widetilde{\mu} = (\mu_u + \mu_d + \mu_s)/3$, which facilitates the
numerical calculations, and $q_f$ represent the electric charge of
each quark flavor.

 The lepton chemical potential follows from the equilibrium reaction
\begin{equation}
e^- \leftrightarrow \mu^- + \nu_e + \bar{\nu}_{\mu} \, ,
\end{equation}
which leads to
\begin{equation}
\mu_e = \mu_{\mu} + \mu_{\nu_e} + \mu_{\bar{\nu}_{\mu}} \, . \label{eq:mue}
\end{equation}
Neutrinos are trapped in the very early stages of the life of a
proto-neutron star, during which it is assumed that the lepton
fraction is kept constant. This can be expressed mathematically as
\begin{eqnarray}
 Y_{Le} = \frac{n_e + n_{\nu_e}}{ n } = \xi ,\nonumber \\ Y_{L\mu} =
 \frac{n_{\mu} + n_{\nu_{\mu}}}{ n } = 0 \, .
\end{eqnarray}
During this phase, the stellar matter is opaque to neutrinos and its
composition is characterized by three independent chemical potentials,
which are $\mu_n$, $\mu_e$, and $\mu_{\nu_e}$. The condition
$Y_{L\mu}=0$ accounts for the fact that no muons are present in the
matter when neutrinos are trapped. The value of $\xi \simeq 0.4$
depends on the efficiency of electron capture reactions during the
initial state of the formation of proto-neutron stars
\cite{Prakash97}.

When the star cools down, the stellar matter becomes transparent to
neutrinos so that $\mu_{\nu_e} = \mu_{\bar{\nu}_{\mu}} = 0$.  In this
case the number of independent chemical potentials is reduced from
three to two, $\mu_n$ and $\mu_e$.

\subsection{Dense matter phase transition and hybrid EoS}

To model the phase equilibrium between hadronic matter and quark
matter, we shall assume that this equilibrium is of first order and
Maxwell-like, that is, the pressure in the mixed quark-hadron phase is
constant. Theoretically the transitions could be Gibbs-like as well,
depending on the surface tension at the hadron-quark interface. The
value of the surface tension is only poorly known. Lattice gauge
calculations, for instance, predict surface tension values in the
range of $0-100$ MeV~fm$^{-2}$ \cite{KAJANTIE1991693}. According to
theoretical studies, surface tensions above around $70$ MeV~fm$^{-2}$
favor the occurrence of a sharp (Maxwell-like) quark-hadron phase
transition rather than a softer Gibbs-like transition
\cite{sotani2011, Yasutake2014}. In this paper, we consider a sharp
Maxwell-like transition.

Given the theoretical models for quark matter and hadronic matter
discussed in Sects.\ \ref{3nPNJL} and \ref{hyper}, we now proceed to
construct models for the hybrid EoS of compact stars. The EoS for
both the hadronic phase and the quark phase is given by the Gibbs relation
\begin{equation}
 \epsilon = - P + T S + \sum_i \mu_i \, n_i \, ,
 \label{eq:EoS}
\end{equation}
where $P=-\Omega$, $S = \frac{\partial P}{\partial T}$ and $n_i =
\frac{\partial P}{\partial \mu_i}$ ($i$ stands for all the particles
of each phase, including leptons).  The lepton contributions to $P$ and
$S$ follow from $\Omega_L$ given by Eq.\ (\ref{eq:leptons}).

\begin{figure}[ht]
\begin{center}
\includegraphics[width=.48\textwidth]{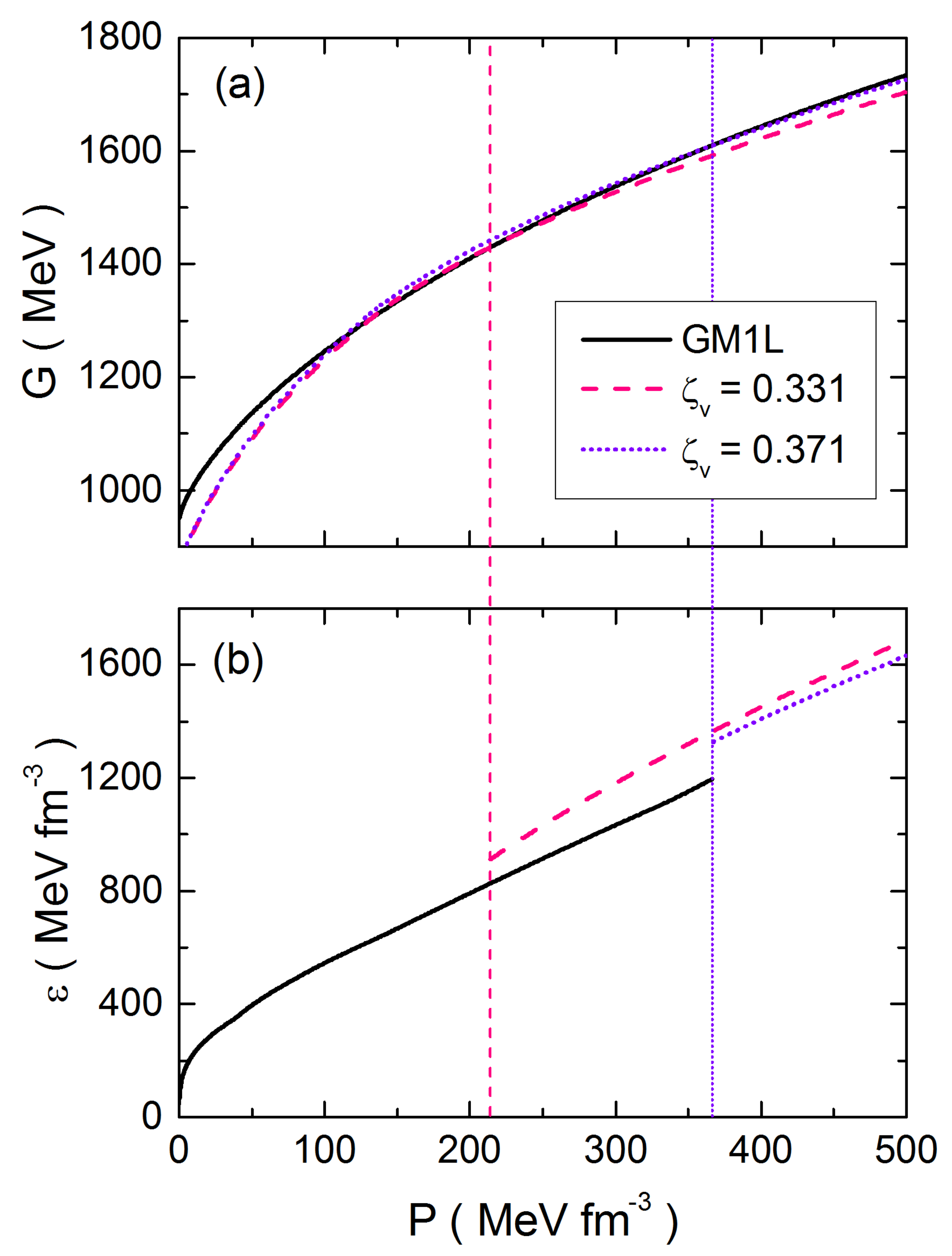}
\caption{(Color online) Panel (a) shows the construction of the EoS
  (at $T=0$) from the Gibbs free energy (per particle), $G$, for the
  hybrid GM1L-3nPNJL parametrization. The solid black line represents
  the hadronic (GM1L) EoS and the red dashed and blue dotted lines are
  the EoSs of the quark (3nPNJL) phase for two values of the vector
  coupling constant, $\mathrm{\zeta_v}$. Panel (b) shows the energy
  density, $\epsilon$, as function of pressure, $P$, for the two
  values of $\mathrm{\zeta_v}$, discussed in the text.}
  \label{Gibbs_gm1l}
\end{center}
\end{figure}

\begin{figure}[ht]
\begin{center}
\includegraphics[width=.48\textwidth]{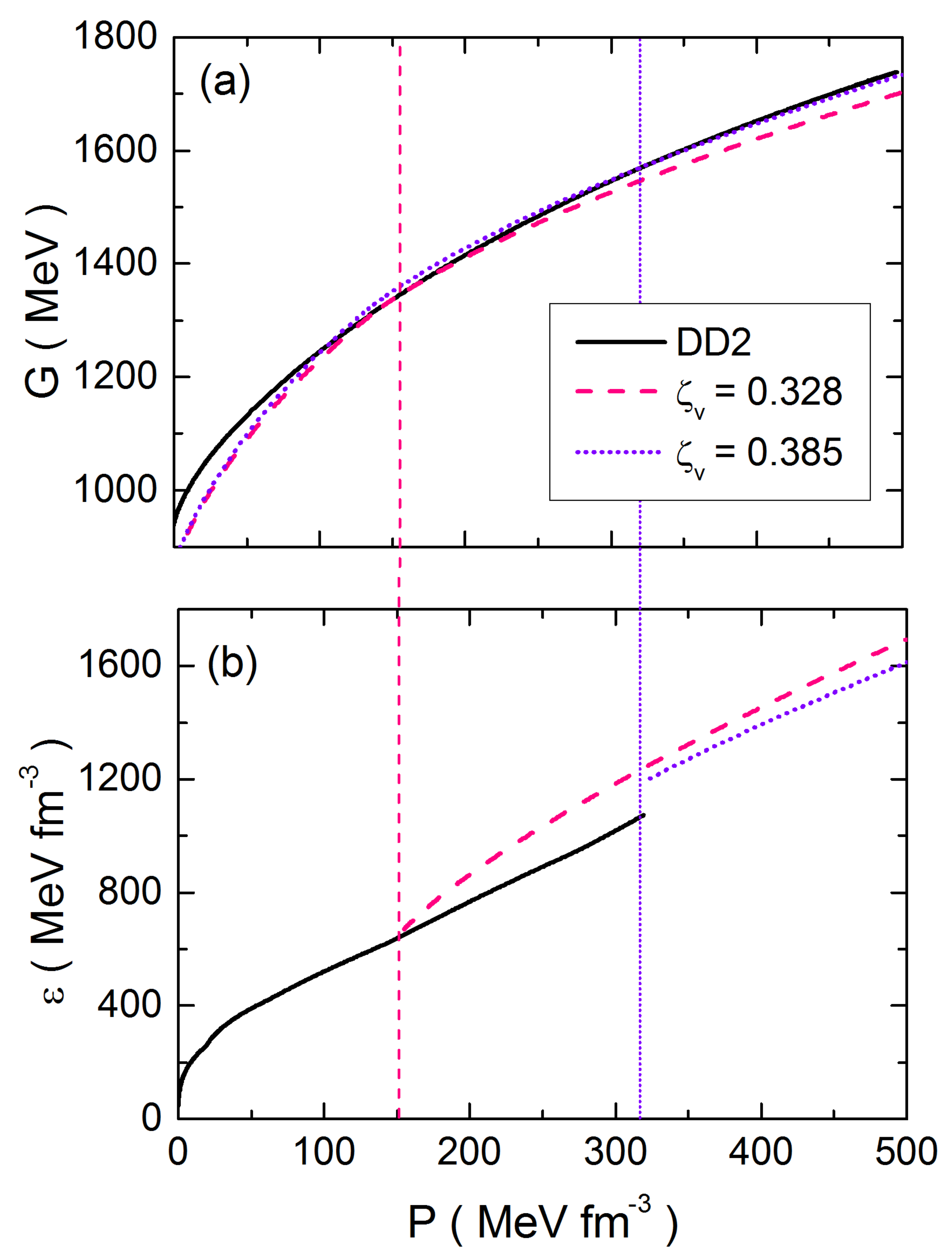}
\caption{(Color online) Panel (a) shows the construction of the EoS
  (at $T=0$) from the Gibbs free energy (per baryon) $G$ for hybrid
  DD2-3nPNJL parametrization. The solid black line represents the
  hadronic (DD2) EoS and the red dashed and blue dotted lines are the
  EoSs for the quark (3nPNJL) phase for two values of the vector
  coupling constant, $\mathrm{\zeta_v}$. Panel (b) shows the energy
  density, $\epsilon$, as function of pressure, $P$, for the two
  values of $\mathrm{\zeta_v}$, discussed in the text.}
  \label{Gibbs_dd2}
\end{center}
\end{figure}

To construct the hadron-quark phase transition we adopt the Gibbs
condition, i.e., the phase transition between both phases
occurs when
\begin{eqnarray}
G_H(P, T) = G_Q (P, T) \, ,
\end{eqnarray}
where $G_{H}$ respectively $G_{Q}$ are the Gibbs free energy per
baryon for the hadronic ($H$) and quark ($Q$) phases at a given
pressure and transition temperature. The Gibbs energy of each phase
($i=H,Q$) is given by
\begin{eqnarray}
G_i (P,T) = \sum_j \frac{n_j}{n} \mu_j \, ,
\end{eqnarray}
where the sum over $j$ is over all the particles present in each
phase. It is important to remark that this is the correct treatment to
calculate a phase transition when different particle species are
present in both phases. In the case of Fig.\ \ref{granpo_rho}, one is
allowed to use both the Gibbs free energy or the chemical potential to
model the phase transition, since there are quarks in both phases. In
contrast, for the hadron-quark phase transition, the particle chemical
potentials in each phase are different so that is becomes necessary to
calculate the Gibbs free energy as a function of pressure to construct
the phase transition \cite{Hempel2013}, as done in Figs.\ \ref{Gibbs_gm1l} and
\ref{Gibbs_dd2} for the GM1L and DD2 parameterizations,
respectively. In these figures, two transitions are visible, the
    first one from quark to hadronic matter at pressures $P \sim 100 -
    150$ MeV/fm$^3$, and the second one from hadronic to quark matter
    at $P \sim 350 - 400$ ~ MeV/fm$^3$. The hadronic and the quark
    matter EoS are very similar and this makes it difficult to
    distinguish between the two phases in the range of the relevant
    pressures, $P \sim 100 - 400$ MeV/fm$^3$. This can be interpreted
    as a masquerade behavior of dense matter, different from
    pure deconfined quark matter \cite{Alford_2005}.
\begin{figure}[ht]
\begin{center}
\includegraphics[width=0.49\textwidth]{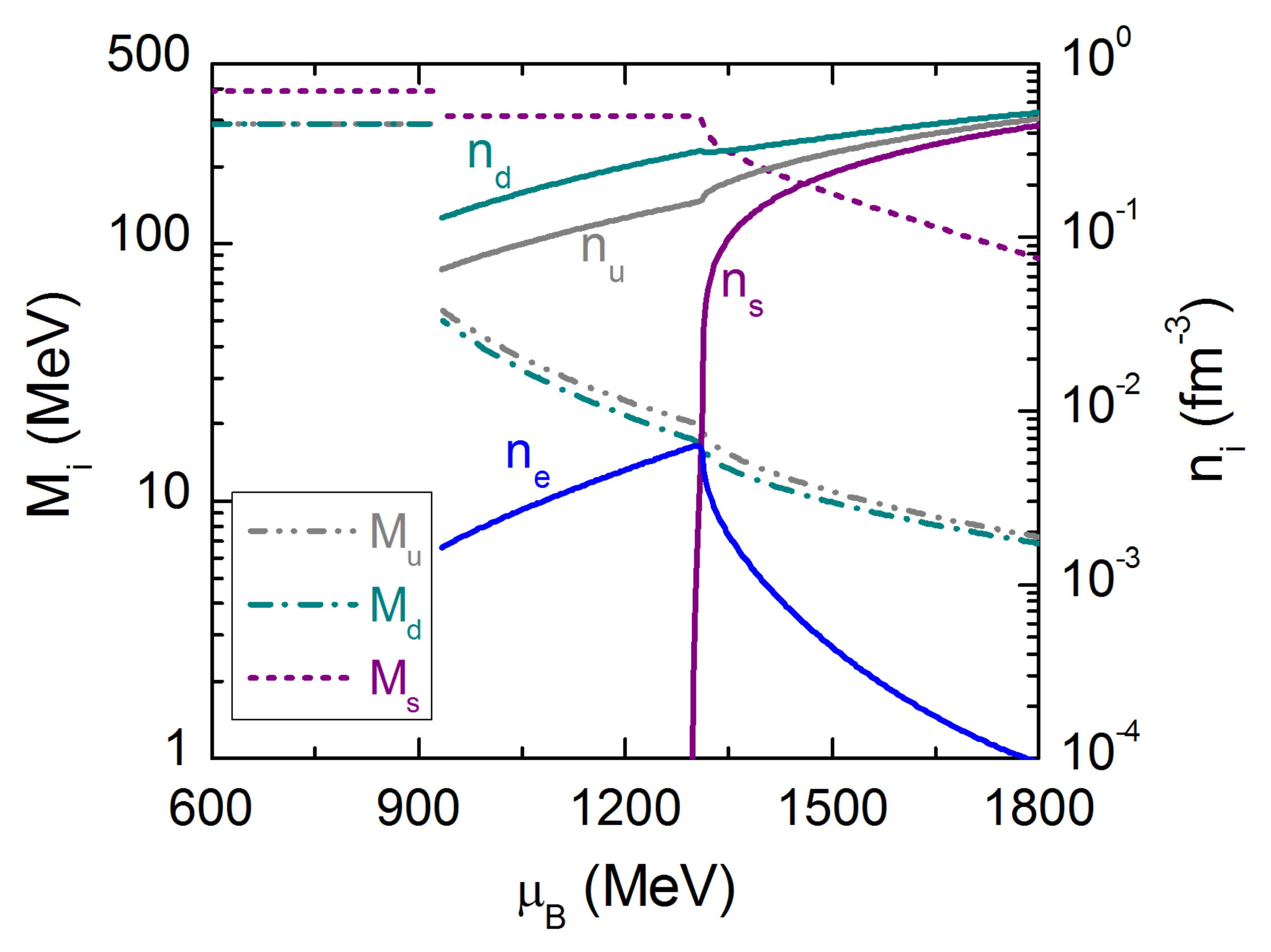}
\caption{(Color online) Dynamical masses, $M_i$, and number
      densities, $n_i$, of \textit{up}, \textit{down} and \textit{strange} quarks as a function of
      baryon chemical potential.  $n_e$ denotes the number density of
      electrons. }
  \label{masses_gibbs}
\end{center}
\end{figure}
 It can be seen from Fig. \ref{masses_gibbs} that a first order
    phase transition occurs at a baryonic chemical potential of $\mu_B
    \sim$ 940 MeV, indicated by the discontinuities in the particle
    number densities and the dynamic quark masses.  For baryon
    chemical potentials between 940 MeV and 1300 MeV we have a phase
    where the chiral quark condensate of the \textit{up} and \textit{down} quarks, $\langle \bar u\, u\rangle\, =\, \langle \bar d\,d\rangle \,\sim \,0$,  while $\langle \bar s\,s\rangle \neq 0$. Such phase exhibits a structure similar to hadronic matter and
    the first quark phase-to-hadron transition is unphysical with
    condensed strange quasi-particles states (the first crossing of
    hadronic and quarks matter curves in Figs.\ \ref{Gibbs_gm1l} and
    \ref{Gibbs_dd2}). This behavior could indicate the existence of a phase which has both aspects of nuclear and quark
    matter (see \cite{McLerran:2007qj, Baym:2017whm}, and references
    therein). Beyond $\mu_B$ $\sim 1300$ MeV and P $\sim 135$
    $\mathrm{MeV/fm^3}$, the strange quarks suffer a crossover
    transition and then deconfine, becoming part of the deconfined
    quark phase used to construct the hybrid EoS.  In this regime \textit{up}
    and \textit{down} quarks could form diquarks and condense in a color
    superconducting state, provided the value of the diquark coupling
    is sufficiently large \cite{Blaschke2009}.

\begin{figure}[ht]
\begin{center}
\includegraphics[width=.48\textwidth]{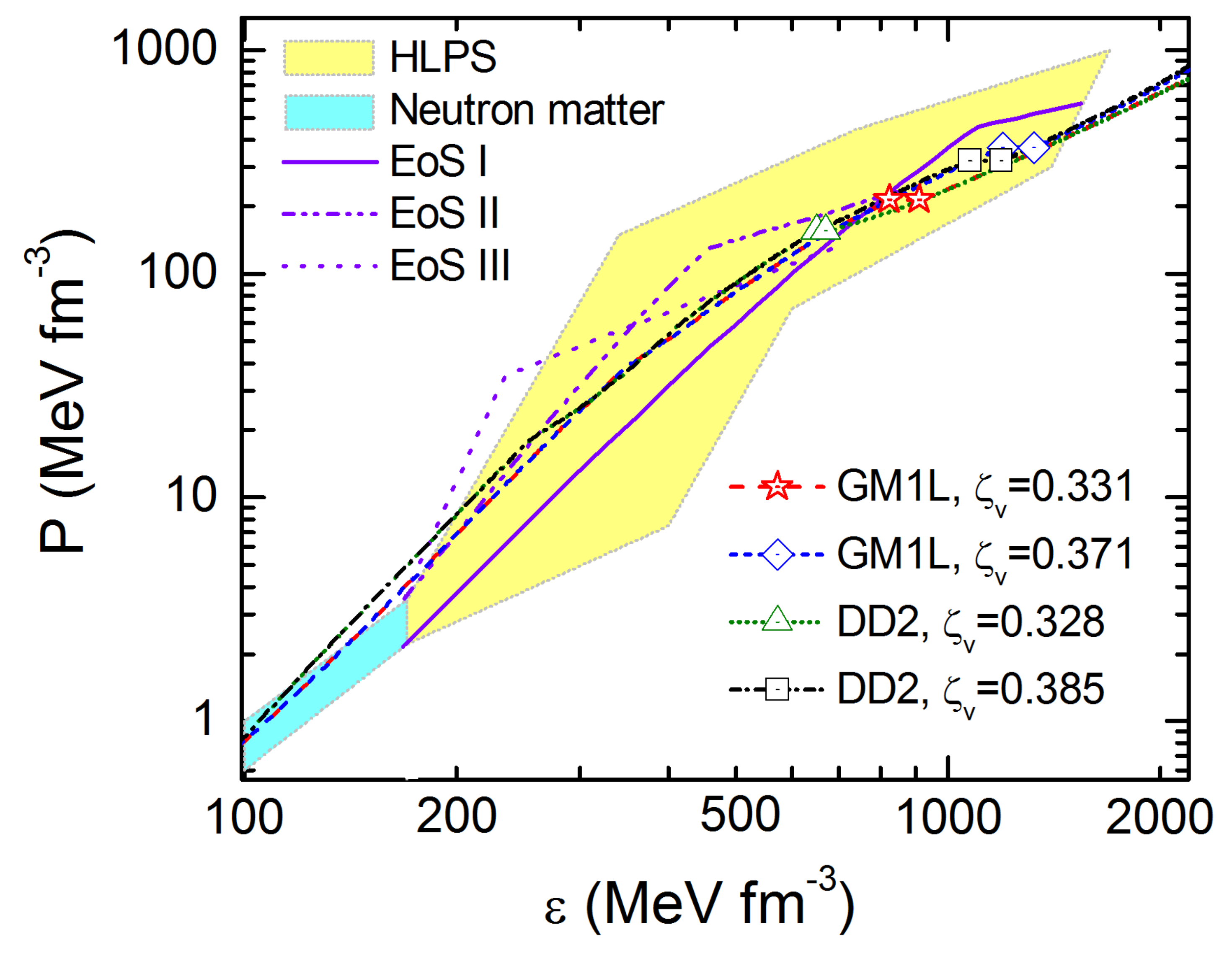}
\caption{(Color online) Comparison of the zero-temperature quark-hybrid
  EoSs of this work (GM1L, DD2) with models (HLPS, Neutron EoS I, EoS
  II, EoS III) from the literature
  \cite{Kruger:2013kua,Hebeler:2013nza,Kurkela:2014vha}. The solid
  dots mark the beginning and the end of the quark-hadron phase for
  our EoSs. The symbol $\mathrm{\zeta_v}$ denotes the vector
  interaction strengths. }
\label{fig:eos_constraints}
\end{center}
\end{figure}

The crossing of the Gibbs energy of the two phases in the $G-P$ plane
defines the phase transition point for a given transition temperature,
$T_\textrm{{trans}}$.  The $2 \, M_\odot$ constraint of PSR
J1614-2230 and PSR J0348+0432
\cite{Demorest:2010bx,Lynch:2012vv,Antoniadis:2013pzd,Arzoumanian_2018}
and the assumption that quark matter exists
in the cores of neutron stars have been used to determine the range of
the vector coupling constant $\mathrm{\zeta_v}$ in the quark
matter. This leads to $0.331<\mathrm{\zeta_v}<0.371$ for GM1L, and
$0.328<\mathrm{\zeta_v}<0.385$ for DD2, where the lower bounds are
determined by the $2\, M_\odot$ constraint and the upper
bounds by the existence of quark matter in the cores of
neutron stars.  It is worth noticing that the density range
  covered by the 3nPNJL model is such that the spinodal region (and
  hence the possible hadronization of deconfined quark matter) is not
  encountered.

\begin{figure}[ht!]
\begin{center}
\includegraphics[width=.48\textwidth]{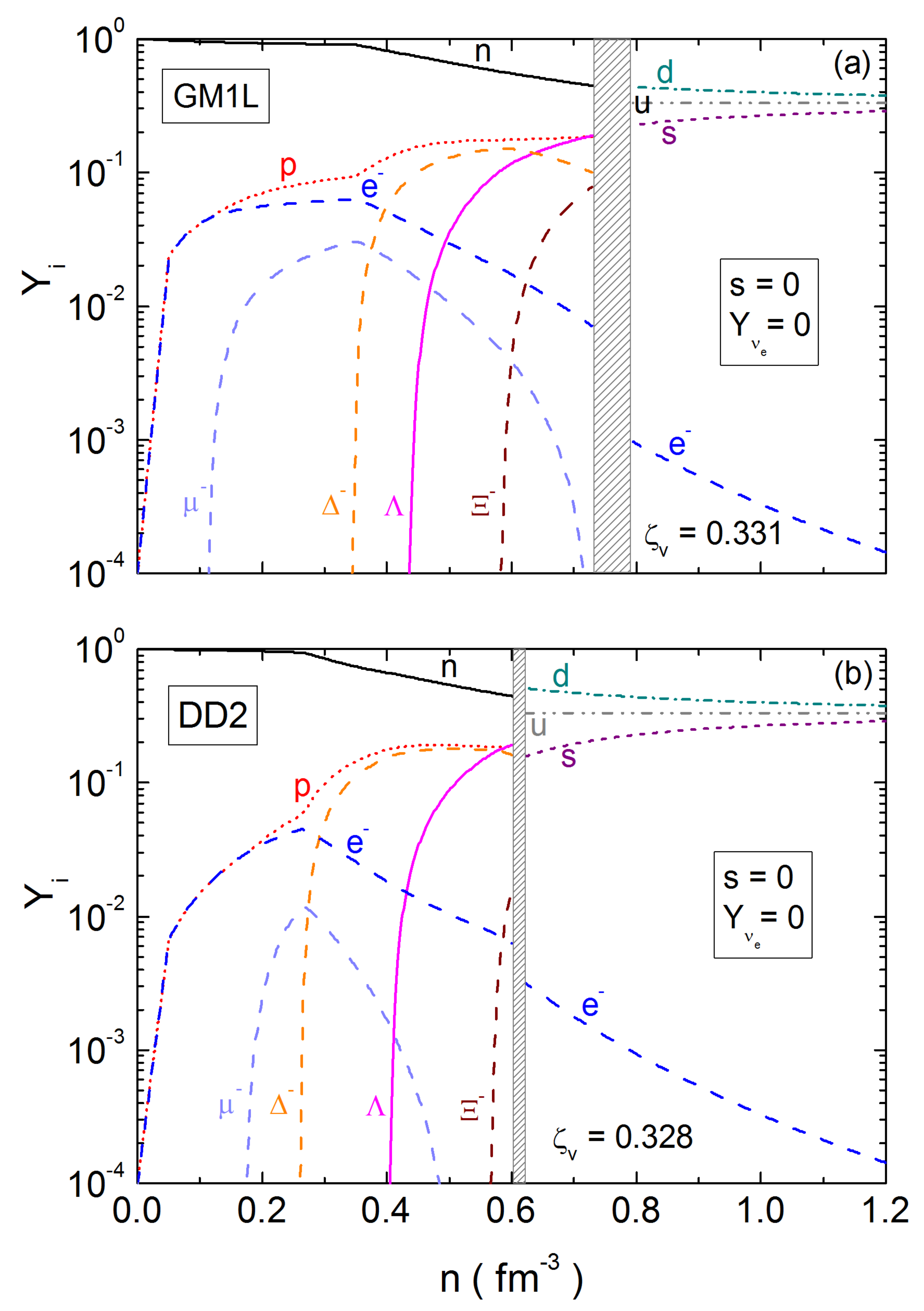}
\caption{(Color online) Particle population of stellar quark-hybrid
  matter at zero temperature as a function of baryonic number
  density. The populations are computed for GM1L-3nPNJL and DD2-3nPNJL
  (see text). $\mathrm{\zeta_v}$ denotes the strength of the vector
  repulson among quarks.}
  \label{part_pop_T0}
\end{center}
\end{figure}

The quark-hybrid EoSs GM1L-3nPNJL and DD2-3nPNJL computed at zero
temperature are compared in Fig.\ \ref{fig:eos_constraints} with
nuclear EoSs suggested in the literature. The curves labeled EoS I,
EoS II, and EoS III are the EoSs determined by Kurkela {\it et
  al.}\ \cite{Kurkela:2014vha}, which are based on an interpolation
between the regimes of low-energy chiral effective field
theory and high-density perturbative QCD. The region labeled HLPS has
been established by Hebeler, Lattimer, Pethick, and Schwenk, and the
area labeled `Neutron matter' shows the equation of state of
low-density neutron matter \cite{Kruger:2013kua,Hebeler:2013nza}.  It
can be seen that the super-dense portions of the hybrid EoSs obtained
in our work are well within these limits.

In Fig.\ \ref{part_pop_T0}, we shown the quark-hadron compositions of
cold neutron stars computed for GM1L-3nPNJL and DD2-3nPNJL.
As expected,  the diversity of particles is significantly reduced at $T=0$.
Even though, the $\Delta^-$-isobar still plays an important role in our
calculations as it reduces the lepton population notably.
As can also be seen, the only
strangeness-carrying hyperons that contribute to the composition are the
$\Lambda$'s and the $\Xi^-$'s, in sharp contrast to the finite $T$ case
(Figs.\ \ref{GM1L_pop_evol} and \ref{DD2_pop_evol}).
A comparison of the GM1L and DD2 populations shows that the particle
abundances are qualitatively similar to each other, and the threshold
densities of the individual particles species are only shifted modestly.

\subsection{Properties of static equilibrium configurations}

To determine the mass-radius relationship for (proto-) neutron stars
we solve the Tolman-Oppenheimer-Volkoff (TOV) equation
\cite{Tolman:1939jz} given by
\begin{equation}
\label{tov1}
\frac{dP}{dr} = -\frac{m(r) \epsilon(r)}{r^2} \frac{ [1 + P(r) /
    \epsilon(r) ] [1 + 4 \pi r^3 P(r) / m(r)]}{1-2 m(r)/r} \,,
\end{equation}
where $P(r)$ and $\epsilon(r)$ are the pressure and energy density at
a radial distance $r$ from the star's center. The gravitational mass
follows from integrating
\begin{equation}
\label{tov2}
\frac{dm}{dr} = 4 \pi r^{2} \, \epsilon(r) \, .
\end{equation}
from $r=0$ to the star's radius, $R$. The latter is defined by $P(R)=0$. The
star's total gravitational mass is thus given by
\begin{equation}
  M_G \equiv 4 \, \pi \int_0^R r^2 \epsilon(r)  dr \, .
\end{equation}
In Sect.\ \ref{sect:PNS_evol} we will discuss stages in the evolution
of proto-neutron stars to neutron star in the gravitational-mass
versus baryon-mass diagram.  The latter is given by
\begin{equation}
  M_B = m_n\int_0^R\frac{4 \pi r^2 n(r)}{[1-2Gm(r)/r]^{1/2}}dr \, ,
\end{equation}
where $m_n = 939$ MeV is the nucleon mass.

\begin{figure}[ht!]
\begin{center}
\includegraphics[width=.48\textwidth]{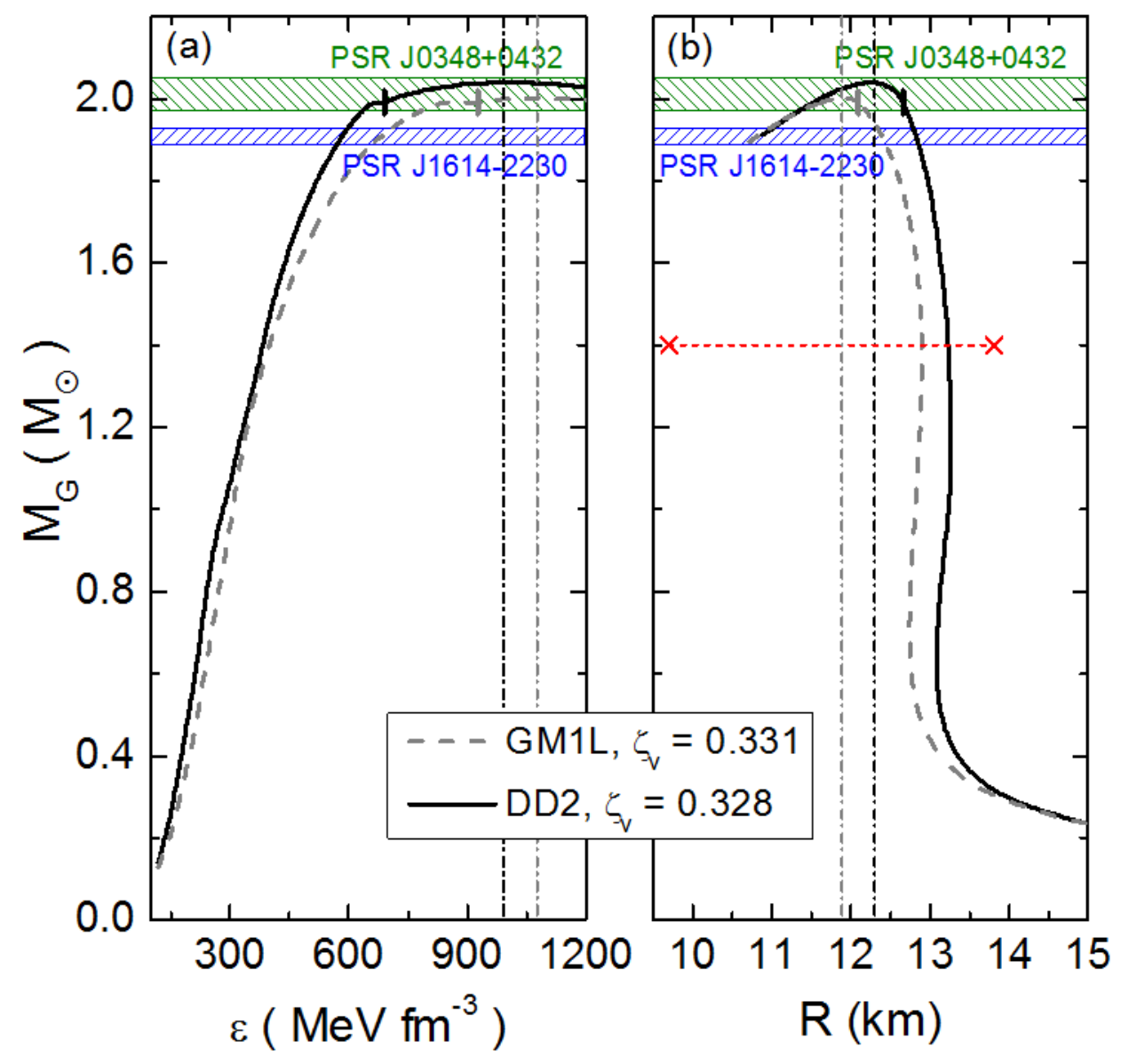}
\caption{(Color online)(a) Gravitational mass as a
    function of central energy density, and (b) gravitational mass as
    a function of stellar radius, for $T = 0$ MeV. The vertical bars
    show the onset of the transition of hadronic matter to quark
    matter. The vertical dash-dotted lines mark the location of the
    maximum-mass star for each EoS. The dashed horizontal line shows
    the estimate of a $1.40 M_\odot$ neutron-star radius derived from
    GW170817 \cite{Bauswein:2017vtn, Fattoyev:2017jql,
    Raithel:2018ncd, Most:2018hfd, Annala:2017llu}.}
  \label{mr_me_T0}
\end{center}
\end{figure}

We first perform the calculations at zero temperature. The results
will be compared with the finite temperature and neutrino trapped
results in Sect.\ \ref{sec:apns}.  Fig. \ \ref{mr_me_T0} shows the
gravitational mass as a function of central energy density as well as
a function of stellar radius for the minimum vector interaction
coupling constants of each hadronic parametrization. The properties of
the maximum-mass stars are summarized in Table \ref{tabla_MR_MB}.
As can be seen, both the pure hadronic EoS as well as the hybrid Eos
lead to maximum-mass neutron stars with fulfill the $2M_{\odot}$ mass
constraint. We also note that the DD2 neutron stars contain a wider
branch of quark-hybrid stars than the GM1L stars, since the DD2 EoS is
stiffer in terms of the Gibbs free energy so that the hadron-quark
phase transition occurs at a lower pressure.

  Color superconductivity (CSC) has not been taken into account in
    this work since a number of problems (such as the diagonalization
    of the Polyakov loop in color space) need to be overcome
    first. However, based on the works carried out in
    Ref.\ \cite{Ruester:2005jc} for a local three-flavor model and in
    Ref.\ \cite{Alvarez-Castillo:2018pve} for a non-local two flavor
    model, one could expect that incorporating CSC into our model will
    shift the onset of the hadron-quark phase transition to lower
    densities, provided, of course, the results of
    \cite{Ruester:2005jc,Alvarez-Castillo:2018pve} have their
    quantitative correspondence in the theoretical model studied in
    this paper.  If so, this would somewhat increase the amount of
    quark matter in the cold neutron stars of our paper.  Their
    maximum masses, however, will not be impacted much since they are
    almost exclusively determined by the hadronic parts of the
    equations of state.  The situation is much harder to assess for
    CSC quark matter at finite temperature (entropy) for conditions
    prevailing in the cores of proto-neutron stars. Chiefly among the
    open issues is the actual size of the gap(s) in the CSC phase
    which, for a given condensation pattern, depend on the density and
    the critical temperature of the CSC phase. Any in-depth
    calculation attempting to address this issue is hampered by the
    fact that the gap(s) is (are) to be computed for quark matter
    constrained by the conditions of color neutrality, electric charge
    neutrality, and chemical equilibrium
    \cite{Blaschke:2005.3flavor}.

\begin{table}[ht!]
\begin{center}
\begin{tabular}{|c|c|c|c|c|}\cline{1-4}
\multicolumn{4}{|c|}{$\mathrm{GM1L}$} \\\hline
& $M_{G}~[M_{\odot}]$ & $M_{B}~[M_{\odot}]$ & $\epsilon_{c}
            ~ [\mathrm{MeV/fm^{3}}]$ \\\hline
$\mathrm{Pure\,\, hadronic}$ & $2.04$ & $2.42$ & $1194.82$ \\
$\mathrm{\zeta_v} = 0.331$ & $2.00$ & $2.36$ & $1077.02$ \\
$\mathrm{\zeta_v} = 0.371$ & $2.04$ & $2.42$ & $1295.79$ \\\hline
\multicolumn{4}{|c|}{$\mathrm{DD2}$} \\\hline
& $M_{G}~ [M_{\odot}]$ & $M_{B}~ [M_{\odot}]$ & $\epsilon_{c}
            ~[\mathrm{MeV/fm^{3}}]$ \\\hline
$\mathrm{Pure\,\, hadronic}$ & $2.11$ & $2.53$ & $1110.68$ \\
$\mathrm{\zeta_v} = 0.328$ & $2.04$ & $2.43$ & $992.88$ \\
$\mathrm{\zeta_v} = 0.385$ & $2.11$ & $2.54$ & $1194.82$ \\\hline
\end{tabular}
\caption{Gravitational mass, $M_G$, and baryon mass, $M_B$, of the
  maximum-mass neutron stars (zero temperature)
  computed for GM1L and DD2. The quantity $\epsilon_c$ denotes the
  stars' cental density.}
\label{tabla_MR_MB}
\end{center}
\end{table}
   \begin{figure}[ht!]
\begin{center}
\includegraphics[width=.48\textwidth]{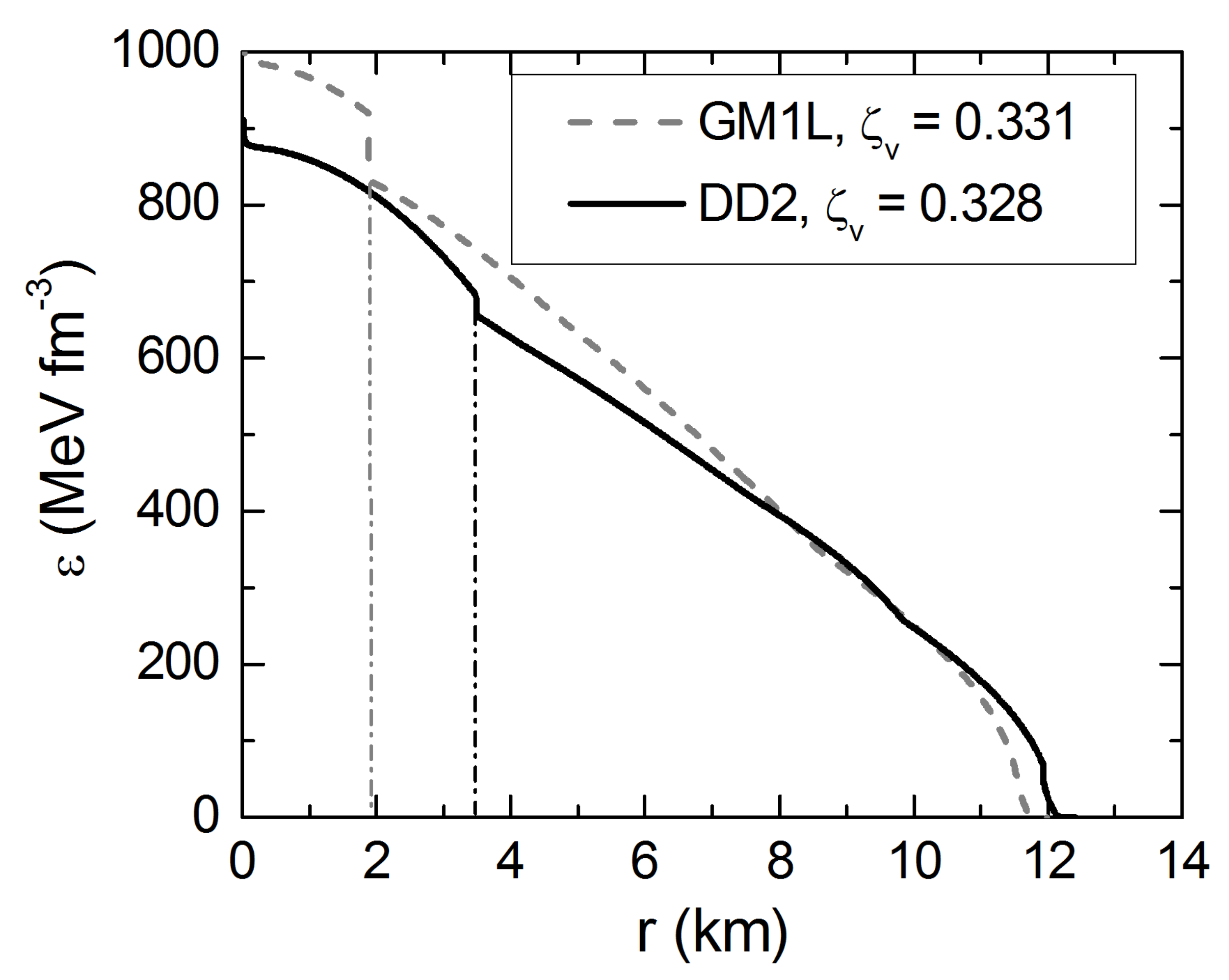}
\caption{(Color online). Energy density as a function of radius
      for the maximum-mass neutron stars shown in
      Fig.\ \ref{mr_me_T0}.  The density discontinuities at around 2
      and 3.5 km (dash-dotted vertical lines) are caused by the
      quark-hadron phase transition in these stars.}
  \label{perfiles-mmax}
\end{center}
\end{figure}

  In Fig.\ \ref{perfiles-mmax} we present energy density profiles
    for the maximum-mass stars shown in Fig.\ \ref{mr_me_T0}. As can
    be seen, these stars contain quark matter cores that are several
    kilometers in size, i.e., $R_{\rm core} \sim 3.5$ km for the DD2
    parameter set and $R_{\rm core} \sim 2$ km for the GM1L.

\section{Application to proto-neutron stars}\label{sec:apns}

\subsection{Finite temperatures and mass-radius relationship}

To study proto-neutron stars we need to extend the EoSs of this work
to finite temperatures. It is known from previous works (see, for
example, \cite{pons}) that proto-neutron stars are nearly isentropic
and not isothermal.  To obtain an isentropic hybrid EoS for the
Maxwell construction, we first compute the hadronic and the quark EoS
for a given transition temperature (i.e., 15 and 30 MeV).  Upon
determining the crossing point of these EoSs in the $G-P$ plane, we
then determine the isentropic hybrid EoS for that transition
temperature.

As already mentioned before, neutrinos play an important role for the
composition of newly formed, hot proto-neutron stars.  For example, it
has been shown in Ref.\ \cite{pons} that during the deleptonization
phase, the stellar core of a proto-neutron star is heated by neutrino
transport (Joule heating), and that the maximum heating occurs just
before the neutrinos escape from the star.  The maximum temperature
reached at this evolutionary stage is around $T \simeq 40-45$
MeV. As a result, different lepton and neutrino fractions at given
entropy values are to be considered when studying different stages in
the evolution of proto-neutron stars to neutron stars. This will be done in
section \ref{sect:PNS_evol} below.

\begin{figure}[ht]
\begin{center}
\includegraphics[width=.47\textwidth]{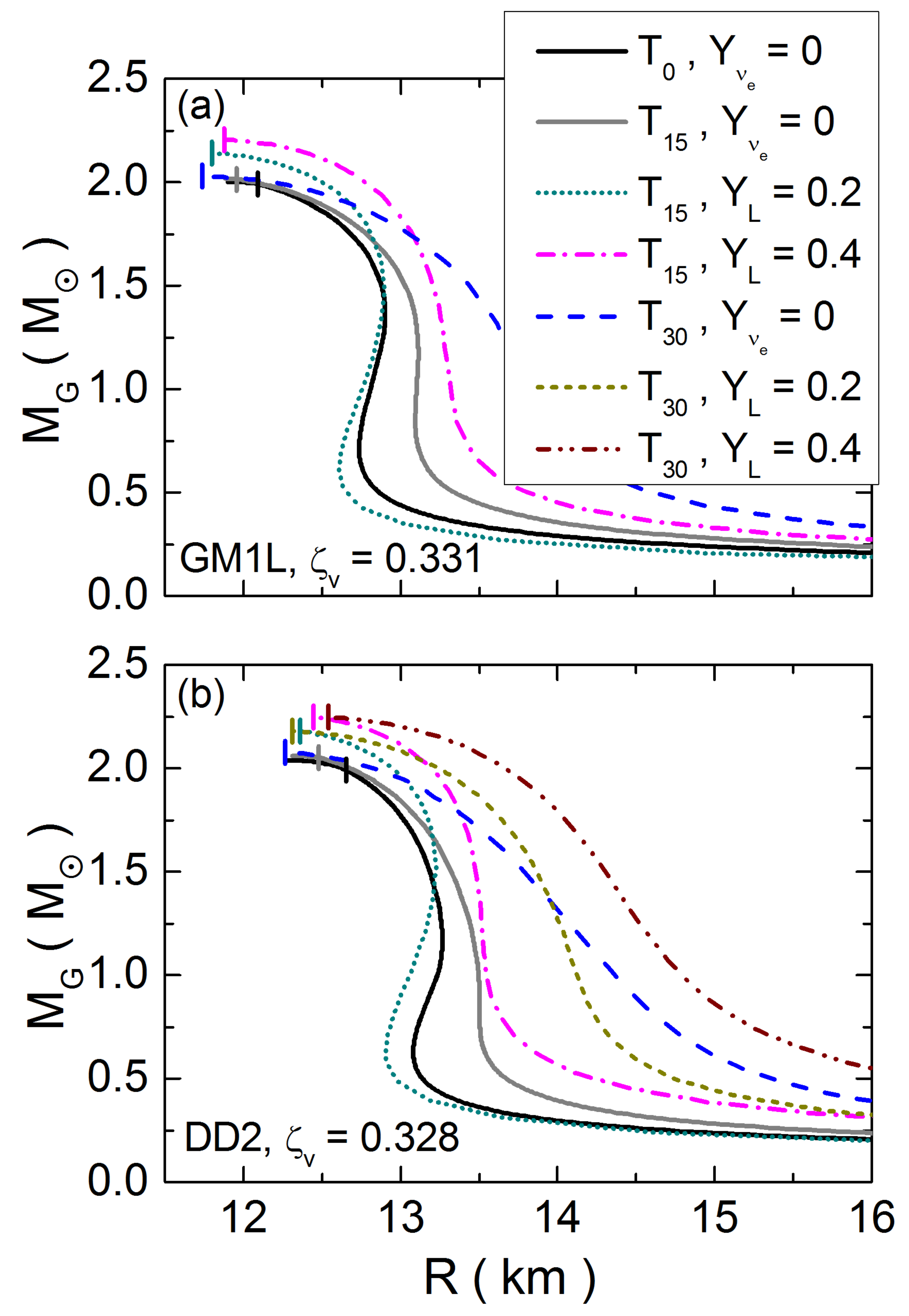}
\caption{(Color online) Gravitational mass, $M_G$, as a function of
  radius, $R$, for quark-hybrid stars at different transition
  temperatures, $T_\textrm{{trans}}$, of the quark-hadron phase
  transition. $Y_L$ denotes the lepton fraction and $Y_{\nu_e}$ the
  neutrino fraction. The vertical bars mark the onset of quark
  deconfinement. With the  exception of the neutrino-less
      ($Y_{\nu_e}=0$) stars with $T_{\rm trans}=0$ (${\rm T_0}$) and $T_{\rm trans}=15$ MeV (${\rm T_{15}}$), this
      transition happens at the maximum-mass peak.}
  \label{MR_TEMP}
\end{center}
\end{figure}

We begin this section by studying the effects of temperature on the
properties of hot stars. For this purpose we have constructed
isentropic EoSs for the parameterizations of this work, choosing
representative proto-neutron star temperatures of $T = 15$ and $30$
MeV.  Depending on the star's evolutionary stage, the presence of
neutrinos is taken into account too (i.e., $Y_{\nu_e} \neq 0$), and
the lepton fractions that we consider are $Y_L = Y_e + Y_{\nu_e} =
0.2$ or 0.4. The mass radius relationships of stars made up of such
matter are shown in Fig.\ \ref{MR_TEMP}.

For the maximum values of the vector interaction for each hadronic
parametrization ($\mathrm{\zeta_v} = 0.371$ and $\mathrm{\zeta_v} =
0.385$) we found that an increase in temperature (with and without
neutrinos) opposes the formation of quark matter in the cores of
stars.  The only stars found to contain quark matter (for these
$\mathrm{\zeta_v}$ values) are the zero-temperature neutron stars. For
the minimum values of the vector interaction the results are
qualitatively similar to the maximum-value case.  Differences concern
primarily the trapping of neutrinos.  For the DD2 parametrization, for
instance, a hybrid EoSs with trapped neutrinos can be constructed up
to $T_\textrm{{trans}} = 30$ MeV (labeled as ${\rm T_{30}}$ in Fig.\ \ref{MR_TEMP}). For the GM1L parametrization,
however, neutrinos are only present in the matter up to
$T_\textrm{{trans}} = 15$ MeV (${\rm T_{15}}$, for higher transition temperatures, the
stars become unstable before the phase transition occurs).

As expected, in Fig.\ \ref{MR_TEMP} it can be observed that the
influence of neutrino trapping in the maximum mass stars is greater
than those originated from a fixed entropy per baryon. As shown for
example in Ref. \cite{Prakash97}, such influence depends sensibly on
the matter composition, in particular, if heavy hadrons (like hyperons
and $\Delta$-isobars) and quarks are taken into account. This behavior is
in sharp contrast to the idealized EoS containing only nucleons and
leptons and no additional softening components, where neutrino
trapping generally reduce the maximum mass.

\subsection{Dense proto-neutron star matter}

\begin{figure}[ht!]
\begin{center}
\includegraphics[width=.48\textwidth]{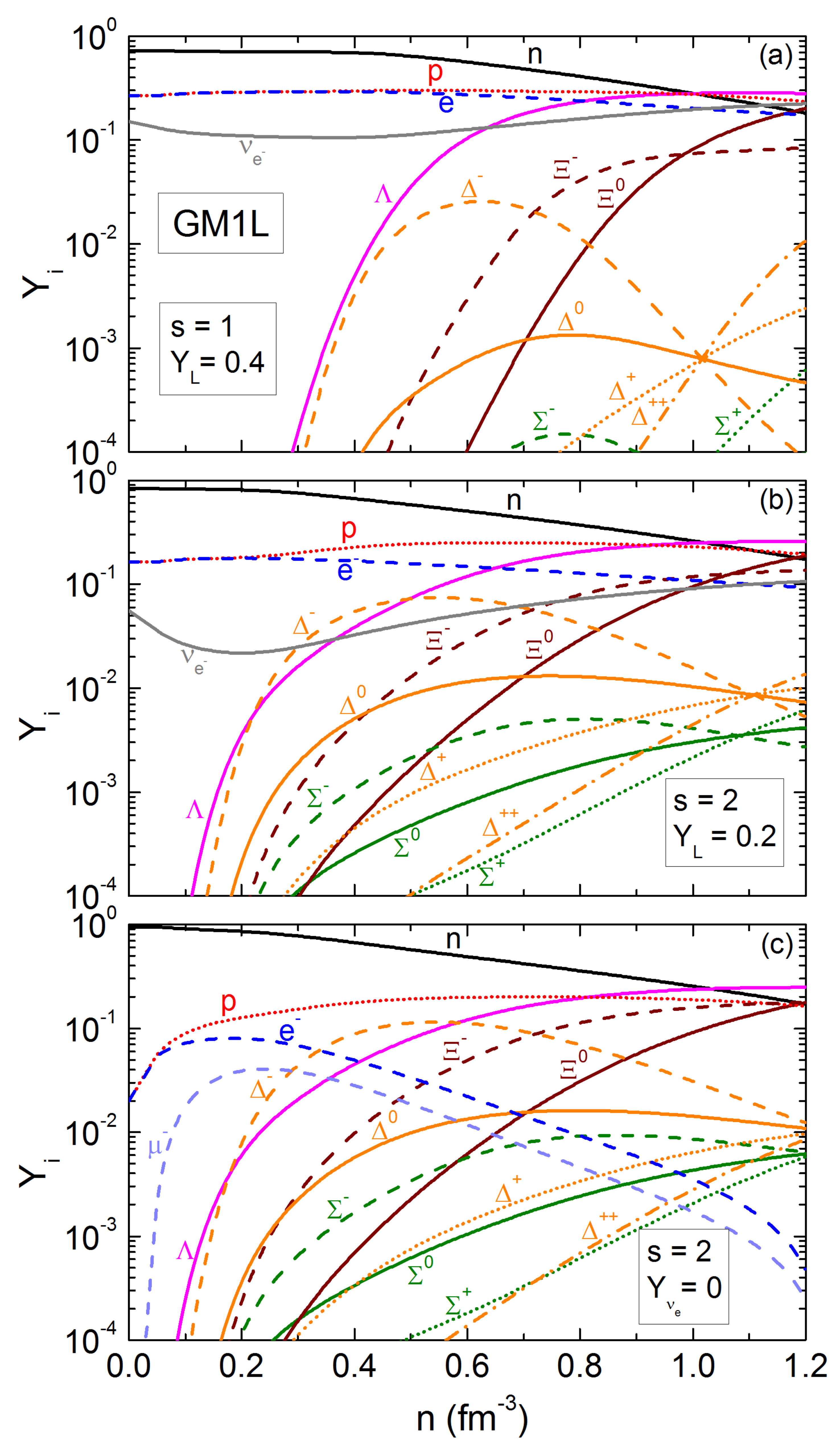}
\caption{(Color online) Particle populations of
    proto-neutron star matter for the GM1L parametrization. The
    compositions correspond to matter in the cores of proto-neutron
    stars at different evolutionary stages characterized by entropy
    per baryon, $s$, and lepton number, $Y_L$.}
  \label{GM1L_pop_evol}
\end{center}
\end{figure}	

\begin{figure}[ht!]
\begin{center}
\includegraphics[width=.48\textwidth]{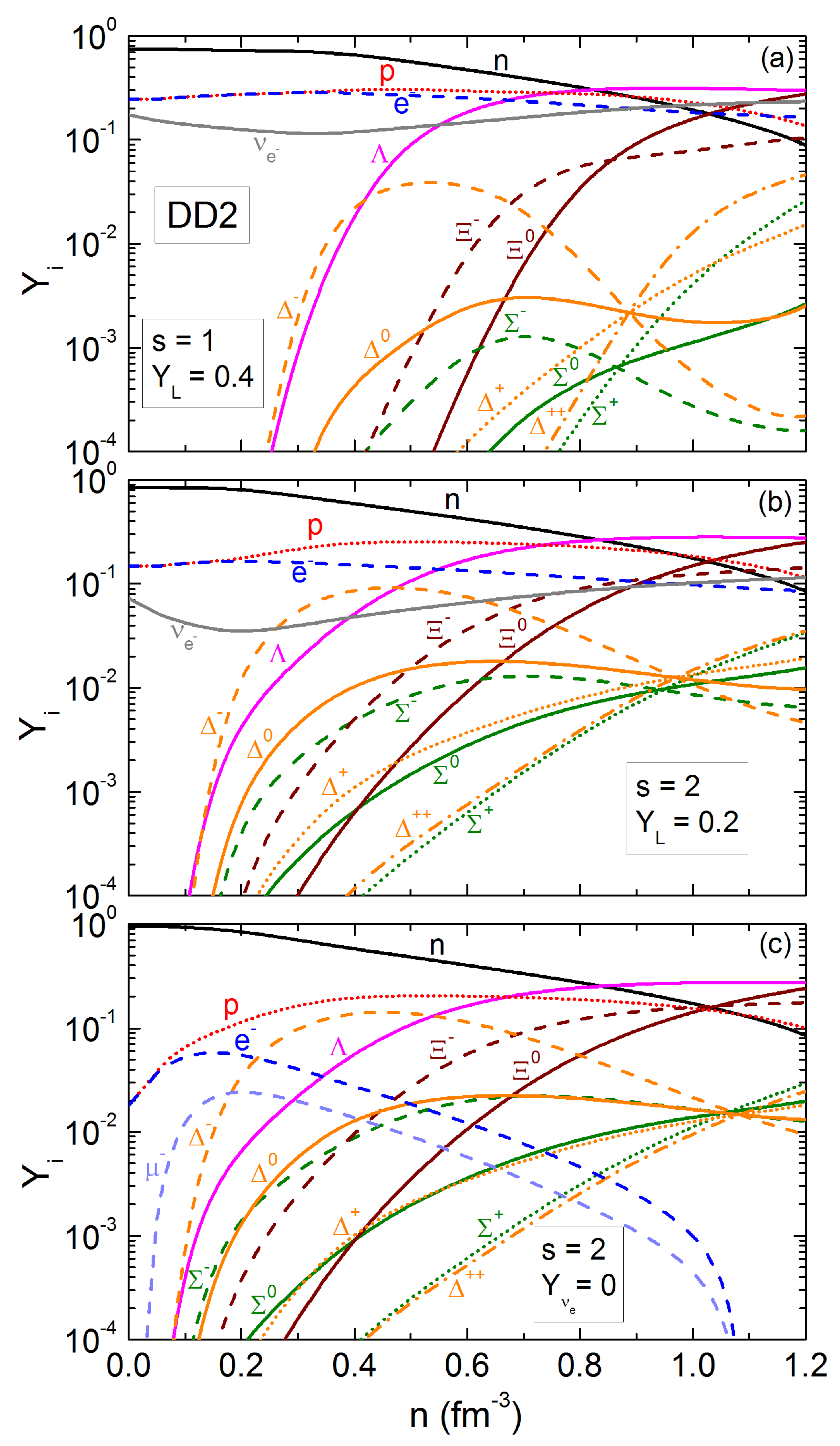}
\caption{(Color online) Same as Fig.\ \ref{GM1L_pop_evol}, but for
  the  DD2 parametrization.}
  \label{DD2_pop_evol}
\end{center}
\end{figure}

Figs.\ \ref{GM1L_pop_evol} and \ref{DD2_pop_evol} show the particle
populations of proto-neutron star matter computed for the hadronic
parametrizations used in this work.  It can be seen that the particle
populations depend sensitively on entropy per baryon, $s=S/n$, and
lepton number, $Y_L$. This is particularly the case for the
$\Delta$-isobar.  The negatively charged state of this
particle are populated first, replacing some of the high-energy
electrons. The other three stages of the $\Delta$-isobar (i.e.,
$\Delta^0$, $\Delta^+$, and $\Delta^{++}$) are successively populated
at densities that are just a few times greater than the nuclear
saturation density. All these stages therefore exist in the cores of
proto-neutron stars, according to our model.  Another striking
difference concerns the high abundance of electrons in matter where
the lepton fraction is non-zero and neutrinos are present (top (a) and
middle (b) panels of Figs.\ \ref{GM1L_pop_evol} and
\ref{DD2_pop_evol}). Because of that, one may speculate that the
electric conductivity of such matter is considerably different from
the electric conductivity of neutrino-free stellar matter (bottom (c)
panels of Figs.\ \ref{GM1L_pop_evol} and \ref{DD2_pop_evol}), where
the presence of muons leads to fewer electrons in the system, and the
increasing $\Delta^-$, $\Xi^-$, and $\Sigma^-$ populations cause a
further reduction of the number of leptons.  Regarding the
strangeness-carrying hyperons, their main contributions come from the
$\Lambda$'s and $\Xi$'s, whose populations grow monotonically with
density, dominating the stellar matter composition at very high
densities.  Other hyperons species are also present, but to a lesser
degree.

\subsection{Stages in the evolution of proto-neutron stars to
neutron stars}
\label{sect:PNS_evol}

In this section, we use the EoSs of this paper to study several stages
in the evolution of proto-neutron stars to neutron stars
\cite{Prakash97}. Shortly after core bounce a proto-neutron star is
hot and lepton rich. The entropy per baryon and lepton fraction of the
matter in the core of such an object change quickly from around $s=1$
and $Y_L=0.4$ to $s=2$ and $Y_L=0.2$. Subsequent core heating and
deleptonization change these values to $s=2$ and $Y_{\nu_e}=0$,
leading to a hot lepton-poor neutron star in less than a minute after
the star's birth \cite{pons}.  After several minutes this hot neutron
star has cooled down to temperatures less than 1 MeV, that is, the
star has become cold.  From then on, the star continues to slowly cool
via neutrino and photon emission until the thermal radiation becomes
too weak to be detectable with x-ray telescopes.

\begin{figure}[htb]
\begin{center}
\includegraphics[width=0.49\textwidth]{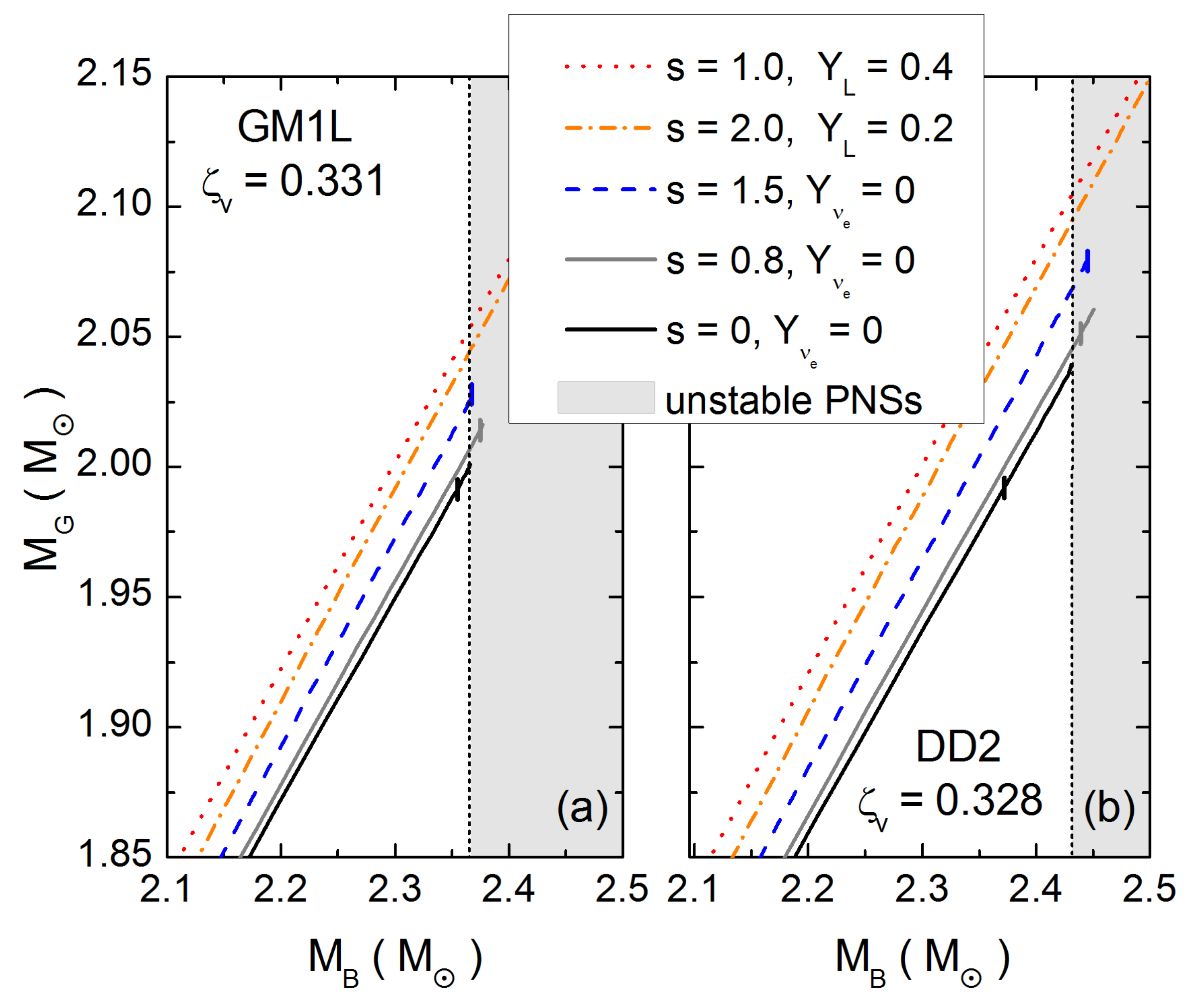}
\caption{(Color online) Gravitational mass versus baryonic
    mass of selected stages (characterized by entropy and lepton
    number) in the evolution of proto-neutron stars to neutron stars,
    computed for the EOSs of this paper.  Each line terminates at the
    maximum-mass star of each stage. The small vertical bars close to
    the maximum-masses mark the onset of the hadron-quark phase
    transition. Only the most massive members of the cold neutron-star
    sequence are found to have pure quark matter in their
    cores. Stars in the shaded region are gravitationally unstable.}
  \label{MGvsMB}
\end{center}
\end{figure}

In Fig.\ \ref{MGvsMB}, we show the gravitational-mass versus
baryon-mass relationship of stars with entropies and lepton numbers
that correspond to the different stages in the evolution of
proto-neutron stars to neutron stars described just above.  Assuming
we are working with isolated stars, the baryonic mass should be a
conserved quantity along the different stages of stellar evolution. As
an example, this condition is represented by a vertical dashed line
passing through the maximum mass cold star in Fig.\ \ref{MGvsMB}.  The
short vertical bars in this figure mark the onset of quark
deconfinement in the cores of these stars. Proto-neutron stars in
their earliest stages of evolution (i.e., $s=1$, $Y_L=0.4$ and $s=2$,
$Y_L=0.2$) are found to be made of pure hadronic matter, no matter how
massive. Once these stars have deleptonized ($Y_{\nu_e}=0$) and their
core entropies have dropped to entropies of $s=1.5$ and 0.8, the
density at quark deconfinement sets in is reached. But this turns out,
for our sample stars, to happen only in stars that are in the
gravitationally unstable region (shaded areas in Fig.\ \ref{MGvsMB}),
where the proto-neutron stars have greater baryonic mass than the
corresponding maximum mass cold star.  The situation is different once
the temperature has dropped to just a few MeV, that is, when these
stars have turned into cold ($s=0$, $Y_{\nu_e}=0$) neutron stars,
which possess pure quark matter in their cores. In Tables
\ref{mass-radius} and \ref{mass-radius2} we show the changing core
compositions of proto-neutron stars as they evolve to the
associated maximum-mass cold stars.

\begin{table}[ht!]
{%
\begin{center}
\begin{tabular}{|l|c|c|l|}\cline{1-4}
\multicolumn{4}{|c|}{GM1L and $\mathrm{\zeta_v} = 0.331$} \\ \hline
\multicolumn{1}{|c|}{Stages} & $M_G~ [M_{\odot}]$ & $R~ {\rm [km]}$
& {Core compositions}
\\ \hline
$s=1.0\,, ~Y_L = 0.4$ & $2.05$ & $12.75$  &Pure hadronic\\
$s=2.0\,, ~Y_L = 0.2$ & $2.04$ & $12.84$  &Pure hadronic\\
$s=1.5\,, ~Y_{\nu_e} = 0$  & $2.02$ & $ 11.94 $ &Pure hadronic \\
$s=0.8\,, ~Y_{\nu_e} = 0$  & $2.01$ & $ 11.97 $ &Pure hadronic \\
$s~\,=0.0\,, ~Y_{\nu_e} = 0$  & $2.00$ & $11.90$   &Quark-Hybrid\\ \hline
\end{tabular}
\caption{Masses, radii, and core compositions of the (proto-) neutron
  stars with conserved baryonic mass $M_B = 2.36 M_{\odot}$, obtained
  for the GM1L parametrization. }
\label{mass-radius}
\end{center}
}
\end{table}

\begin{table}[ht!]
{%
\begin{center}
\begin{tabular}{|l|c|c|l|}\cline{1-4}
\multicolumn{4}{|c|}{DD2 and $\mathrm{\zeta_v} = 0.328$} \\ \hline
\multicolumn{1}{|c|}{Stages} & $M_G ~[M_{\odot}]$ & $R~ {\rm [km]} $
&{Core compositions}
\\ \hline
$s=1.0\,, ~Y_L = 0.4$ & $2.10$ & $13.09$ &Pure hadronic \\
$s=2.0\,, ~Y_L = 0.2$ & $2.09$ & $13.15$ &Pure hadronic \\
$s=1.5\,, ~Y_{\nu_e} = 0$  & $2.07$ & $ 12.37 $ &Pure hadronic \\
$s=0.8\,, ~Y_{\nu_e} = 0$  & $2.05$ & $ 12.49 $ &Pure hadronic \\
$s~\,=0.0\,, ~Y_{\nu_e} = 0$  & $2.04$ & $12.27$  &Quark-Hybrid\\ \hline
\end{tabular}
\caption{Same as Table \protect{\ref{mass-radius}}, but for the DD2
  parametrization and a conserved baryonic mass of $M_B = 2.43
      M_{\odot}$.}
\label{mass-radius2}
\end{center}
}
\end{table}

It has been proposed \cite{Prakash97,Brown:1993jz} that the unstable
proto-neutron stars mentioned above will collapse to black
holes. Moreover, it has been shown in
Ref.\ \cite{Prakash97,Vidana:2002rg} that the collapse to a black hole
could also be related to the presence of hyperons, $\Delta$-isobars,
and/or quarks in the stellar matter, since the hot neutrino-trapped
matter is capable of supporting more massive objects than cold stellar
matter.

\subsection{Tidal deformability of neutron stars}

The tidal deformability of neutron stars is an important parameter for
gravitational-wave (GW) astronomy as it determines the pre-merger GW
signal in NS-NS merger events.  To linear order, the tidal
deformability, $\lambda$ is given by
\begin{equation}
\lambda = - \frac{\mathcal{E} _{ab}}{Q_{ab}} \, , \nonumber
\end{equation}
where $\mathcal{E} _{ab}$ is the applied external field and $Q_{ab}$
the induced mass-quadrupole moment. $\lambda$ is related to
the dimensionless tidal Love number, $k_2$, associated with $\ell =2$
perturbations,
\begin{equation}
\lambda = \frac{2}{3}k_2R^5 \, , \nonumber
\end{equation}
where $R$ denotes the stellar radius.  The dimensionless tidal
deformability, $\Lambda$, can then be calculated as
\begin{equation}
\Lambda = \lambda/M^5 \, ,
\end{equation}
where $M$ denotes the star's gravitational mass.
The tidal Love number can be written in terms of the stellar
compactness, $\beta = M/R$, as
\begin{align}
k_2 \, = \, & \bigg\{ \frac{8}{5}\beta^5(1-2\beta)^2 \Big[
  2+2\beta(\eta-1)-\eta \Big] \bigg\}  \nonumber \\ & \times
\bigg\{ 2\beta \Big[6-3\eta+3\beta(5\eta-8) \Big]  \\ & + 4\beta^3
\Big[13-11\eta+\beta(3\eta-2)+2\beta^2(\eta+1) \Big]  \nonumber \\ &
+ 3(1-2\beta)^2 \Big[2-\eta+2\beta(\eta-1) \Big]\ln(1-2\beta)
\bigg\}^{-1} \nonumber \, ,
\end{align}
with $\eta = \eta(r=R)$.  $\eta(r)$ is the solution of
\begin{eqnarray} \label{eqz}
r\frac{{\rm d}\eta}{{\rm d}r} & + \eta(r)^2 + \eta(r){\rm e}^{\lambda
  (r)}\left[1+4\pi r^2 \left[P(r) + \epsilon (r)\right]\right] \nonumber \\& + r^2
\Xi(r) = 0,
\end{eqnarray}
where
\begin{eqnarray}
\Xi(r) &=& 4\pi{\rm e}^{\lambda (r)}\left[5\epsilon (r) + 9P(r) +
  \frac{\epsilon (r) + P(r)}{{\rm d}P/{\rm
      d}\epsilon}\right]\nonumber\\ & -& 6\frac{{\rm e}^{\lambda
    (r)}}{r^2} - \left(\frac{{\rm d}\nu (r)}{{\rm
    d}r}\right)^2. \nonumber
\end{eqnarray}
\begin{figure}[htb]
\begin{center}
\includegraphics[width=0.49\textwidth]{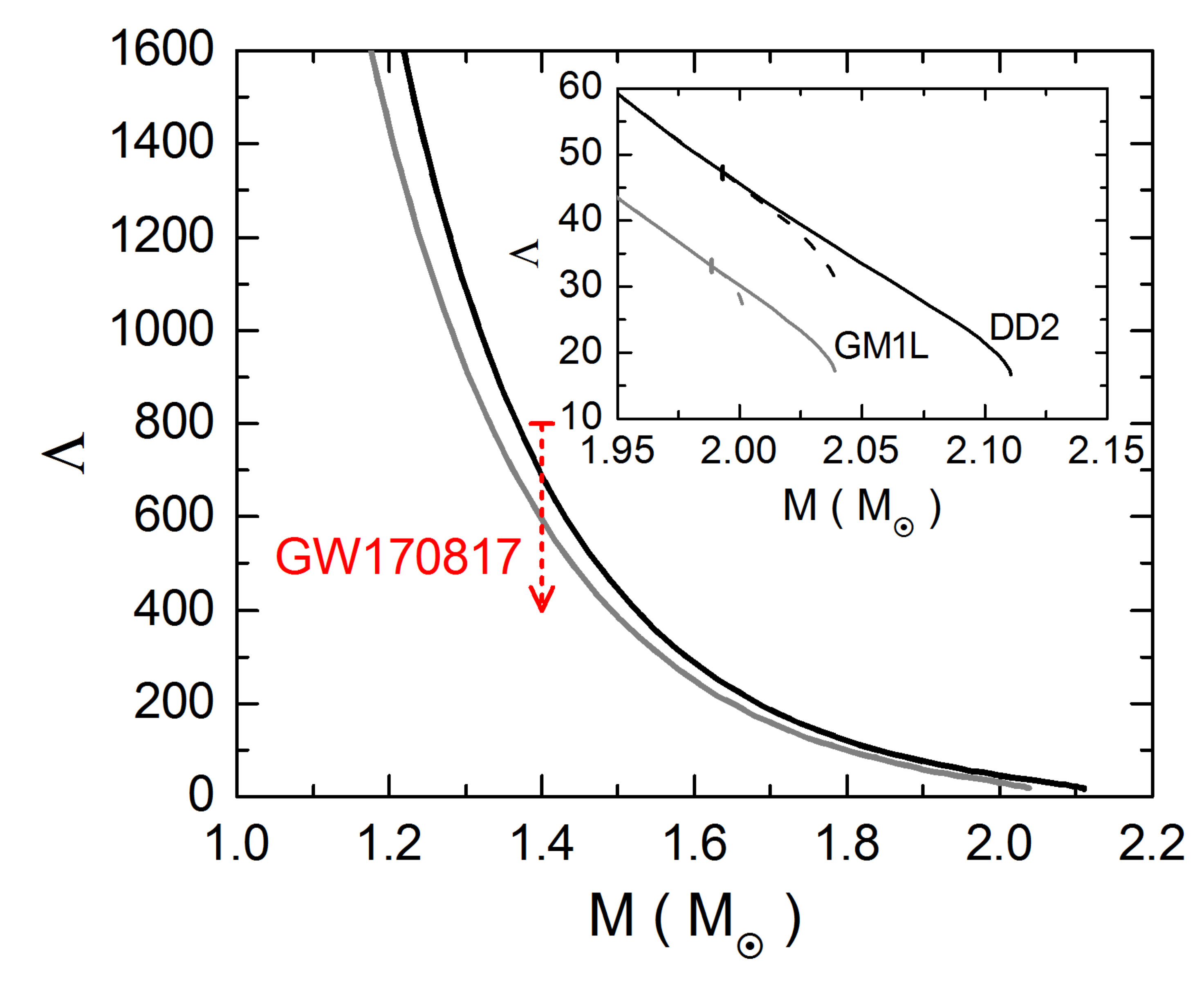}
\caption{(Color online) Tidal deformability versus gravitational
      mass of pure hadronic stars (solid lines). Hybrid branches are
      indicated by dashed lines. The small vertical bars on each curve
      mark the onset of the phase transition from hadronic to quark
      matter.  Each line terminates at the maximum-mass star. The red
      arrow shows the constraint on $\Lambda$ imposed by the analysis
      of the data of GW170817.}
  \label{tidal_fig}
\end{center}
\end{figure}
Equation~(\ref{eqz}) it to be solved simultaneously with the TOV
equation for the boundary condition $\eta(0)=2$.

When an EoS with a sharp discontinuity at a radius $r=r_{\rm d}$ is
used to describe the matter in the interior of a compact object, the
additional junction condition
\begin{equation}
\eta(r_{\rm d}^+) - \eta(r_{\rm d}^-) = \frac{4\pi r_{\rm d}^3
  \left[\epsilon (r_{\rm d}^+) - \epsilon (r_{\rm
      d}^-)\right]}{m(r_{\rm d})} \nonumber
\end{equation}
is to be imposed \cite{2018arXiv181010967H}.

The data analysis of GW170817 puts constrains on the dimensionless
tidal deformability of a $1.4 M_\odot$ star which is given by $\Lambda
_{1.4} \le 800$ (see, \cite{Most:2018, Raithel:2018,
  Abbott:2018,JPGreview} and references therein.)

In Fig.\ \ref{tidal_fig} we present the dimensionless tidal
deformability as a function of gravitational mass for the cold hybrid
stars studied in this work. We also present, for completeness, the
results of purely hadronic neutron stars. Due to the high value of the
transition pressure, the discrepancies are only noticeable for the
high mass objects, being $\sim 10\%$ for the GM1L case and $\sim 20\%$
for the hadronic EoS DD2. The red arrow shows the limit imposed on
$\Lambda$ by the analysis of the data from GW170817. As can be seen,
our results are in agreement with the observational constraint.

\section{Summary and conclusions}\label{summary}

This paper had two main objectives. The first objective was to
investigate the phase diagram of quark matter using the non-local
3-flavor NJL model coupled to the Polyakov loop.  In particular, we
studied the possible existence of a spinodal region in the QCD phase
diagram and determined the temperature and chemical potential of the
critical end point (CEP).

The peaks of the chiral susceptibility of light quarks were used to
determine the crossover phase transition (critical points) in the
phase diagram. For the first-order transition, the spinodal lines have
been determined from the vanishing of the speed-of-sound. As shown in
\cite{Fukushima:2008wg,Contrera:2012wj}, the location of the CEP along
the phase transition line depends on the vector-to-scalar interaction
strength, $\mathrm{\zeta_v}$. We found that considering the vector
interactions shrinks the metastable region in the phase diagram,
renders quark matter less compressible, and shifts the first-order
phase transition to higher chemical potentials.

The second main objective of this paper was to investigate the
quark-hadron composition of baryonic matter at zero as well as
non-zero temperature.  This is of great topical interest for the
analysis and interpretation of neutron star merger events such as
GW170817. With this in mind, we determined the composition of
proto-neutron star matter for entropies and lepton fractions that are
typical of such matter. These compositions were used to delineate the
evolution of proto-neutron stars to neutron stars in the baryon-mass
versus gravitational-mass diagram.

For the treatment of hadronic matter, we used the DDRMF model which
takes into account density-dependent meson-baryon coupling constants.
Vector meson-hyperon coupling constants were chosen according to the
SU(3) ESC08 model, while the scalar meson-hyperon coupling constants
were fitted to empirical hypernuclear potentials. This coupling scheme
leads to hadronic EoSs (labeled GM1L and DD2) which satisfy the $2 \,
M_{\odot}$ constraint as well as the constraint on  neutron star
radii derived from the gravitational-wave event GW170817.

The hadron-quark phase transition was treated as a Maxwell
construction, which leads to a sharp hadron-quark interface.  The $2
\, M_\odot$ constraint of PSR J1614-2230 and PSR J0348+0432 and the
assumption that quark matter exists in the cores of (cold) neutron
stars were used to determine the range of the vector coupling constant
$\mathrm{\zeta_v}$ in quark matter. This lead to
$0.331<\mathrm{\zeta_v}<0.371$ for GM1L, and
$0.328<\mathrm{\zeta_v}<0.385$ for DD2, where the lower bounds follow
from the $2\, M_\odot$ constraint and the upper bounds from
the existence of quark matter in the cores of neutron stars.

The compositions and EoSs of hybrid stars were computed at zero as well as
finite temperature,  entropies $0 \leq s \leq 2$, lepton numbers
$0 \leq Y_L \leq 0.4$,  with and without neutrinos.
The EoSs were then used to delineate the evolution
of proto-neutron stars to neutron stars in the baryon-mass versus
gravitational-mass diagram.  We found that the hybrid-DD2 EoS with
$\mathrm{\zeta_v} = 0.328$ allows for the existence of hybrid stars up
to $T_\textrm{{trans}}$= 30 MeV while the hybrid-GM1L EoS with
$\mathrm{\zeta_v} = 0.328$ leads to hybrid configurations with
critical temperatures less than $T_\textrm{{trans}}$ = 15 MeV.  Based
on the dense matter models of this work, quark matter existing (by
construction) in cold neutron stars, would neither be present in hot
neutron stars nor in proto-neutron stars. The situation is drastically
different for  hyperons and  $\Delta$-isobars, which are found to
exist very abundantly in proto-neutron star matter.

In closing, we mention that the data provided by gravitational-wave
detectors such as LIGO and VIRGO have the potential to shed light on
whether or not hybridization and/or quark deconfinement occurs in the
cores of neutron stars.  Of particular interest in this context is the
tidal deformability of neutron stars which depends strongly on the
nuclear EoS. As discussed in \cite{Abbott:2018,JPGreview} (and
references therein), the tidal deformability determined for the
colliding neutron stars that lead to the gravitational-wave event
GW170817 could provide stringent limits on the existence of quark
matter in the interiors of neutron stars. The tidal deformability
expresses by how much neutron stars are deformed by tidal forces
shortly before they collide. This deformation induces a change in the
gravitational potential, which, in turn, leads to characteristic
changes in the gravitational-wave signal emitted during the collision.
The determination of the tidal deformability, therefore, opens up a
new and exciting window into the inner workings of neutron stars.  The
hope is that the upcoming data collecting runs with Advanced LIGO and
Advanced Virgo will provide exciting new insight into the
deformability of neutron stars and thus the EoS of super-dense matter
itself.

\section*{Acknowledgments}

The authors thank J. Randrup and G. Lugones for
discussions and comments during the preparation of this
manuscript. In addition, the authors thank the anonymous referee for his/her constructive comments, which substantailly helped
improving the original manuscript. This work is supported through the U.S. National Science
Foundation under Grant PHY-1714068. G. M., M. O., G. A. C. and
    I. F. R-S thank CONICET and UNLP for financial support under
grants PIP-0714 and G140, G157, X824.  G. A. C. is thankful for
hospitality extended to him at the San Diego State University and for
the support from the CONICET-NSF joint research project titled
\textit{``Structure and properties on neutron star cores''}.

\bibliography{Malfatti}

\end{document}